\documentclass[showpacs,amssymb,floatfix,pre,aps,notitlepage,preprint,nofootinbib]{revtex4-2}
\usepackage{bm}        % for math
\usepackage{amssymb}   % for math
\usepackage{cancel}

\usepackage{soul}% http://ctan.org/pkg/soul

\usepackage{mathtools}

\usepackage{amssymb,amsfonts,bm}
\usepackage{graphics,graphicx}
\usepackage{colordvi,amsmath,epsfig,color}
\usepackage[activeacute,english]{babel}
\usepackage{appendix}
\usepackage[mathscr]{euscript}
\usepackage{amsmath}
\usepackage{xcolor}
\usepackage{physics}
\usepackage{enumerate}% http://ctan.org/pkg/enumerate

\usepackage{subcaption}
\usepackage{hyperref}
\usepackage{soul}

\newcommand{\un}{\widehat{\bm\sigma}}

\definecolor{darkgreen}{rgb}{0,0.6,0.0}

\begin{document}

\title{Self-diffusion in confined systems}
%\input author_list.tex       % D0 authors (remove the first 3 lines
                             % of this file prior to submission, they
                             % contain a time stamp for the authorlist)
                             % (includes institutions and visitors)

\author{M.  Mayo}
\affiliation{F\'{\i}sica Te\'{o}rica, Universidad de Sevilla,
Apartado de Correos 1065, E-41080, Sevilla, Spain}

\author{M. I. García de Soria}
\affiliation{F\'{\i}sica Te\'{o}rica, Universidad de Sevilla,
Apartado de Correos 1065, E-41080, Sevilla, Spain}

\author{P. Maynar}
\affiliation{F\'{\i}sica Te\'{o}rica, Universidad de Sevilla,
Apartado de Correos 1065, E-41080, Sevilla, Spain}

\author{J. J. Brey}
\affiliation{F\'{\i}sica Te\'{o}rica, Universidad de Sevilla,
  Apartado de Correos 1065, E-41080, Sevilla, Spain}
\affiliation{Institute for Theoretical and Computational
  Physics. Facultad de Ciencias. Universidad de Granada, E-18071,
  Granada, Spain}

\date{\today}

\begin{abstract}
The self-diffusion process of a hard sphere fluid confined by two
parallel plates separated by a distance on the order of the particle
diameter is studied. The starting point is a closed kinetic 
equation for the distribution function that takes into account the
effects of the confinement and that is valid in the low-density
limit. From it, the Boltzmann-Lorentz equation that describes the
dynamics of some tagged particles when the whole system is in
equilibrium is derived. An equation that
describes the diffusion in the directions parallel to the walls is deduced
by applying the Zwanzig-Mori projection technique to the
Boltzmann-Lorentz equation, obtaining an explicit expression for the
self-diffusion coefficient that depends on
the height of the system. A very good agreement between its
theoretical prediction and Molecular
Dynamics simulation results is obtained for the whole range of
heights. 
\end{abstract}

\maketitle

%%%%%%%%%%%%%%%%% Introduction %%%%%%%%%%%%
\section{Introduction}
\label{sec:intro}

Diffusion in fluids is one of the widest studied process in the
context of Statistical
Mechanics. Starting from stochastic methods \cite{vanKampen} or by
kinetic theory \cite{resibois1977classical}, how a particle diffuses is well
understood, at least in the low-density limit. For finite densities, the
appearance of long-time tails in the velocity correlation function
makes the situation more delicate \cite{aw70} but, still, kinetic
theory has provided a microscopic understanding of the phenomena
\cite{c93} for simple models such as hard spheres. In this case, the
obtained theoretical predictions compare reasonably well with 
Molecular Dynamics (MD) simulation results and with experiments for
hard sphere-like colloidal suspensions \cite{cvs98, tahd17}. For a general
fluid, there are also some general results to be remarked 
such as the relation between the scaled self diffusivity and the excess
entropy \cite{d96}, that can be understood in terms of
mode coupling theories \cite{sag01, sag04}.

When the system is confined over regions with a width of the order of the diameter
of the particles, the situation is much more complex and there is no microscopic theory that allows us to understand diffusion as well as in the non-confined case. We will consider situations where the size of the particles is much larger or smaller than that of the particles making up the
walls, so that the last ones can be considered to be flat. We will
assume that the particles move freely between collisions,
corresponding to situations where the influence of the curvature and motion of the walls on the
movement of the particles is felt on a much larger scale than the mean free path, i.e. large persistence length compared to the mean free path (note that, in
rarefied conditions, the dynamics would be dominated by wall
effects). It is also assumed that there is not an interstitial medium
or its effect on the movement of the particles is negligible over
distances of the order of the mean free path.  Under these conditions, 
one of the simplest models to carry out the study is an ensemble of elastic
hard particles confined by two fixed parallel walls separated a distance of
the order of the particles' diameter. In fact, in the last years a
kinetic theory for such a system has been formulated when the system
is ultraconfined, i.e., the separation between the plates is smaller
than twice the diameter of the particles, so that the particles can
not jump over each other. In the low-density limit, the one-particle
distribution function obeys a Boltzmann-like kinetic equation in which
the collisional contribution is modified with respect to the non-confined
case in order to take into account the effects of the confinement
(for example due to the constraints induced by the walls, not all the orientations of the collisions
are possible) \cite{bmg16, bgm17, mbgm22}. The
derivation was heuristic, following Boltzmann arguments. Moreover, an
$H$-theorem has been proved, so that equilibrium is always 
reached in the long-time limit. MD simulations
have also shown that the kinetic equation 
describes very well the evolution to equilibrium in simple situations where the
system is spatially homogeneous (in the direction parallel to the walls) but the temperature parameters in the vertical
and horizontal directions are different. The corresponding
Boltzmann-Lorentz equation was formulated in \cite{bgm20} from which
the diffusion equation was derived by a modification of the
Chapman-Enskog method. An explicit expression for the
diffusion coefficient as a function of the height of the system was
obtained, that agrees very well with MD simulation 
results. 

Very recently, the kinetic equation has been derived by a density
expansion of the Liouville equation \cite{bgm24}. In
fact, the derivation brings to 
the fore that it is not essential that the distance between the plates
be smaller than twice the diameter of the particles, but of the same
order. Here we will extend the analysis performed in 
\cite{bgm20} to wider systems. This is, actually, the objective of the
paper: to formulate the corresponding Boltzmann-Lorentz equation for elastic
hard spheres confined by two parallel plates separated a distance of
the order of the diameter of the particles, in order to study
self-diffusion in the direction parallel to the plates when the system
is in equilibrium. In the analysis performed in the ultraconfined
case \cite{bgm20}, it was possible to understand how the diffusion
properties vary when approaching a pure 2-dimensional (2d) system. In
the present case, we learn how the diffusion properties change from
a quasi-2d (q2d) system to a pure 3d system. 

The paper is organized as follows. In the following section the
microscopic model is introduced while, in Sec. \ref{sec:KT}, the
kinetic equation for the distribution function integrated in the
direction perpendicular to 
the plates is formulated. Some useful mathematical properties of the
collisional operator are discussed and an $H$-theorem is proved. In
Sec. \ref{sec:bl}, the Boltzmann-Lorentz equation that describes the
dynamics of some tagged particles immersed in a bath in equilibrium is
studied and the corresponding diffusion equation is derived following
standard projection methods \cite{zwanzig1960, zwanzig2001,
  mori1965}. An explicit formula for the diffusion coefficient is
obtained as a function of the separation of the plates. Comparison  with MD simulation resultsis made in Sec. \ref{sec:sim}. Very good agreement is found. Finally, a discussion is presented in
Sec. \ref{sec:conc}, together with our conclusions and some
perspectives. Details of the calculations are presented in the
appendices. 

%%%%%%%%%%%%%%%%% The model %%%%%%%%%%%%
\section{The model}
\label{sec:model}

\begin{figure}[t]
\centering 
\includegraphics[scale=0.2]{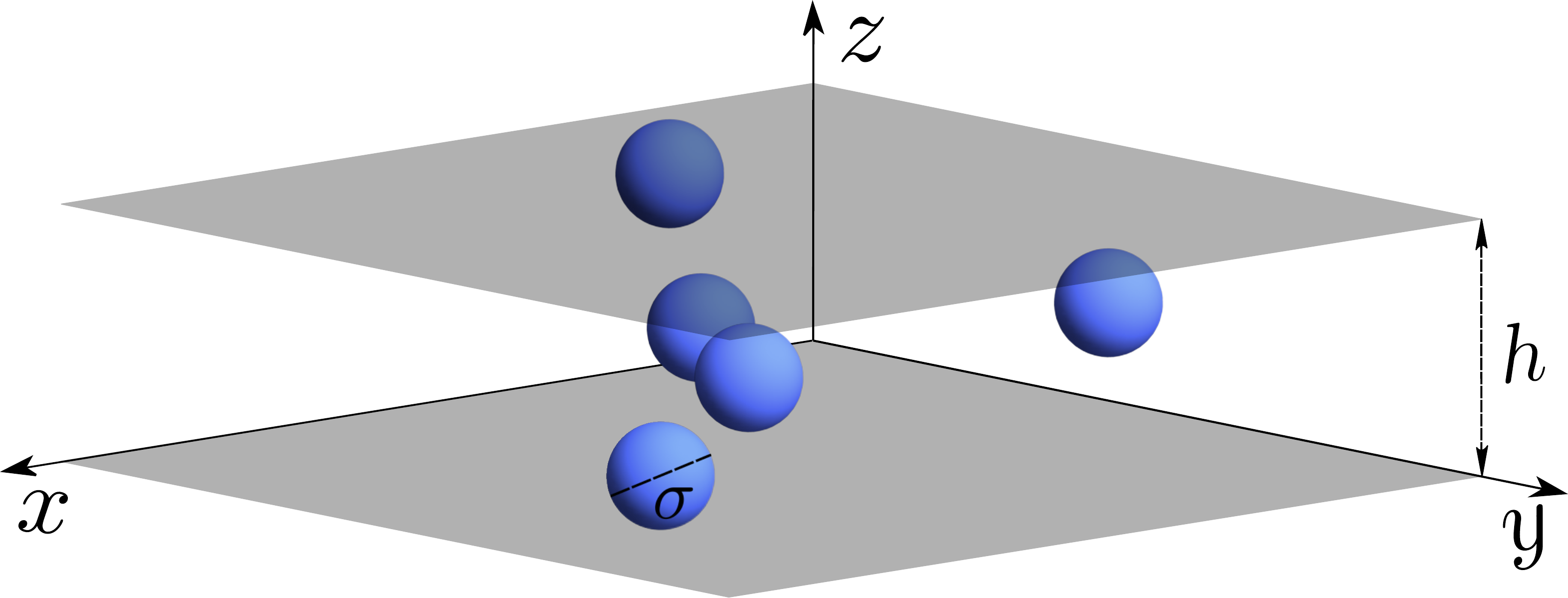}
\caption{Sketch of the system. The particles are hard spheres of diameter $\sigma$ confined between two parallel plates separated a distance  $h\sim \sigma$. }
\label{fig:1}
\end{figure}

Let us consider a three dimensional gas modeled as $N$ elastic hard
spheres of the same mass $m$ and diameter $\sigma$. The gas is
confined between two square shape parallel walls of  area $A$, separated a distance
$h$. The $z$ axis is taken in the direction perpendicular to the
walls and periodic boundary conditions are applied in the $xy$ directions. It is assumed that the height of the system is of the same
order as the diameter of the particles, i.e., $h\sim \sigma$, while
the size in the parallel direction to walls is much larger than the
mean free path. Fig.~\ref{fig:1} shows an scheme of the
system. Particles move freely between binary encounters that are
modeled as instantaneous elastic collisions. When a particle with
velocity $\bm{v}$ collides with another particle with velocity
$\bm{v}_1$, the post-collisional velocities of the particles,
$\bm{v}'$ and $\bm{v}_1'$, are given by the following equations 
\begin{subequations}
\begin{align}
{\bm v}^{\prime} &\equiv  b_{{\widehat{\bm\sigma}}} {\bm v} = {\bm
  v}+ \left( {\bm g} \cdot \widehat{\bm
  \sigma} \right) \widehat{\bm \sigma}, \label{2.1}
  \\
{\bm v}^{\prime}_{1} &\equiv b_{\widehat{\bm\sigma}} {\bm v}_{1} = {\bm v}_{1}- \left( {\bm g} \cdot \widehat{\bm \sigma} \right) \widehat{\bm \sigma}. \label{2.2}
\end{align} 
\end{subequations}
Here, ${\bm g} \equiv {\bm v}_{1} - {\bm v}$ represents the relative
velocity between particles before the collision, and
$\widehat{\bm\sigma}$ is a unit vector directed along the line joining
the centers of the particles at contact, pointing away from the
particle with the initial velocity $\bm{v}$. Total momentum and energy
are conserved in each collision. We also introduce the operator
$b_{\widehat{\bm\sigma}}$, which transforms pre-collisional velocities
into post-collisional velocities. The walls are located at $z=0$ and
$z=h$. Reflecting walls are considered, so that when a particle is
about to collide with one of the walls—meaning it has a positive
$z$-component of velocity, $v_z>0$, to collide with the top wall, or a
negative $z$-component, $v_z<0$, to collide with the bottom wall—the
velocity after the collision, $\bm{v}''$ , conserves the component
parallel to the wall, where its perpendicular component is inverted,
i.e.,  
\begin{align} \label{eq:reflect_walls}
\bm{v}'' =  b_{W}\bm{v} = \bm{v}-2 v_z\bm{\hat{e}_z}.
\end{align}
 $ \bm{\hat{e}_z}$ is the unitary vector in the direction of the
 $z$-axes (perpendicular to the walls pointing from the bottom to the
 top wall) and $b_W$ is the operator that changes the velocity vector
 before the collision with the wall into the velocity after the
 collision.

\section{Kinetic Theory description}
\label{sec:KT}

 We will consider the system in the dilute regime, so that a kinetic
 theory description is expected to be valid. The state of the system
 is given by the one-particle distribution function,
 $f(\bm{r},\bm{v},t)$, defined as usual in kinetic theory as the mean
 density of particles in phase space. In  Ref.~\cite{bgm24} a kinetic
 equation for 
\begin{equation} \label{p-pdf}
 \overline{f}\left(\bm{r}_{\|}, \bm{v}, t\right) \equiv \frac{1}{h-\sigma} \int_{\sigma / 2}^{h-\sigma / 2} \dd z f(\bm{r}, \bm{v}, t),
\end{equation}
where we have introduced the parallel component of a vector as
$\bm{a}_{\|} \equiv a_{x} \bm{e}_{x}+a_{y} \bm{e}_{y}$, was
systematically derived by a
density expansion of the Liouville equation. Assuming that the initial
distribution, $f(\bm{r}, \bm{v}, 0)$, is mirror symmetric with respect
$z=h/2$, and neglecting the recollisions induced by the walls (two particles that have collided close to the wall, can recollide after one of them has collided with the wall), to first
order in the density, we found that 
$\overline{f}$ obeys \cite{bgm24erratum} 
\begin{equation} \label{eq:bteq2d}
\left(\frac{\partial}{\partial t}+\bm{v}_{\parallel} \cdot
  \nabla_\parallel \right) \overline{f}\left(\bm{r}_{\|}, \bm{v},
  t\right)= \mathcal{J}_h[\overline{f}|\overline{f}], 
\end{equation}
 where  the
 \textit{collision operator} $\mathcal{J}_h[\overline{f}|\overline{f}]$ is defined as 
\begin{align} \label{eq:col_op}
\mathcal{J}_h[\overline{f}|\overline{f}]=\frac{ \sigma^{2}}{h-\sigma}
  \int_{\sigma / 2}^{h-\sigma / 2} \dd z \int   \dd \bm{v}_1
  \underset{\Omega_h(z) }{ \int} \dd \un|\bm{g} \vdot \un|
  \Theta(-\bm{g} \vdot \un) (b_{{\widehat{\bm\sigma}}}  -  1)
  \overline{f}\left(\bm{r}_{\|}, \bm{v}_1, t\right)
  \overline{f}\left(\bm{r}_{\|}, \bm{v}, t\right). 
\end{align}
$\Theta$ is the Heaviside step function and  $\Omega_h(z)$ are
the allowed solid angles of collisions between two particles, being
the particle of velocity $\bm{v}$ at $\bm{r}\equiv
(r_\parallel,z)$. The restrictions on the integral over $\un$ arise because not all the orientations of a
collision between two particles are possible due to the presence of
the walls.

\begin{figure}[t!]
\begin{center}
\includegraphics[scale=0.5]{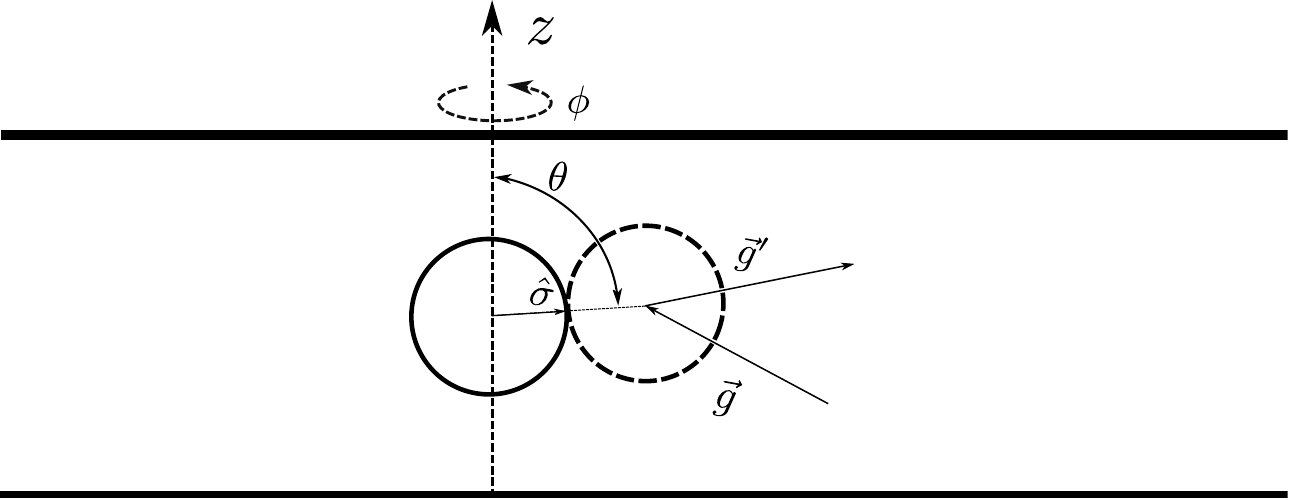} 
\end{center}
\caption{
Sketch of the collision between two particles. The tagged particle is
represented by a solid line circle, while the colliding particle is
represented by a dashed line circle. The vectors $\bm{g}$ and
$\bm{g}^\prime$ are the relative velocities before and after the
collision respectively. The spherical coordinates in which
$\widehat{\bm{\sigma}}$ is expressed are also shown.} 
\label{fig:2}
\end{figure}

Eq.~\eqref{eq:bteq2d} shows that, as in the non-confined Boltzmann
equation, the 
dynamics of $\overline{f}$ has the two typical contributions:
free-streaming and collisions. The independence of the distribution function from $z$ derives from the large number of wall collisions occurring typically during a mean free time. In contrast, the state of the system is
given in terms of a distribution function that only depends on the
horizontal spatial coordinates, although the velocity is still 3d. In
fact, Eq.~\eqref{eq:bteq2d} 
coincides with the Eqs. derived heuristically in \cite{bmg16, bgm17},
when the $z$ dependence of $f$ is not considered. Let us note that in
Ref.~\cite{bgm24} only the case $\sigma < h \leq 2 \sigma $ was
considered. Nonetheless, the arguments are also valid for $h>2\sigma$.  Note that the parametrization of $\Omega_h(z)$
depends strongly on 
$h$ and we have to distinguish  three regions: q2d ($\sigma < h \leq
2\sigma$), intermediate height ($2\sigma \leq h \leq 3 \sigma$) and
wide  ($h\geq 3 \sigma$) systems. Let us also notice that
$\widehat{\bm \sigma}$ will be expressed in spherical coordinates,
being $\theta$ and $\phi$ the polar and azimuthal angles respectively
(see Fig.~\ref{fig:2}).

%%% Q2d systems %%

\setul{5pt}{1.0pt}% 5pt below contents
\begin{itemize}
\item \ul{$\sigma < h \leq 2 \sigma$}
\end{itemize}

\begin{figure}[t!]
\begin{center}
\includegraphics[scale=0.5]{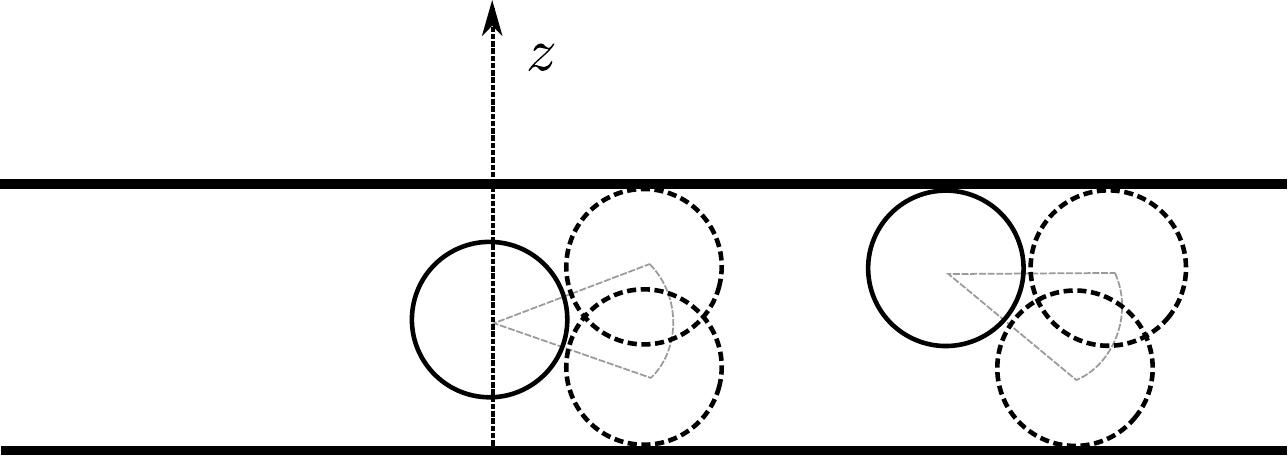} 
\end{center}
\caption{Diagram of the possible solid angles for two different
  heights of the tagged particle (represented by a solid line circle)
  for $\sigma < h  \leq 2 \sigma $. The dashed line circles represent
  particles colliding with the tagged particle with the maximum and
  minimum value of $\theta$. It can be seen that, wherever the
  tagged particle is, the two walls restrict the orientation of the
  possible collisions. } 
\label{fig:3}
\end{figure}

In this case, wherever the tagged particle is, the two walls restrict
the possible orientations of the collision vector $\un$. In Fig.~\ref{fig:3} a
diagram of the possible solid angles for two different $z$-coordinate of the
tagged particle is  represented. These systems are characterized by
the fact that  the particles cannot pass over each other. In
Fig.~\ref{fig:3}, we can see the allowed directions of collisions are
always influenced by the presence of the walls. The solid angles can
be parametrized as follows 
\begin{align}
\label{eq:q2dpar} \Omega_h(z) \equiv 
\begin{aligned}
           \{ (\theta, \phi)|\;\theta\in (\pi/2-b_2(z,h), \pi/2+b_1(z)), \phi\in
           (0,2\pi) \} ,\; & \hbox{ for } \sigma/2 < z< h-  \sigma/2,
    \end{aligned}
\end{align}
with $b_1(z)$ and $b_2(z,h)$, 
\begin{align} \label{eq:b1_b2}
b_1(z) = \sin ^{-1} \left( \frac{ z- \sigma/2}{\sigma} \right), \;
b_2(z,h) = \sin ^{-1} \left( \frac{h- \sigma/2- z}{\sigma} \right).
\end{align}

%%% Itermediate  systems %%

\begin{itemize}
\item \ul{$2\sigma \leq h \leq 3 \sigma$}
\end{itemize}
\begin{figure}[t!]
\begin{center}
\includegraphics[scale=0.5]{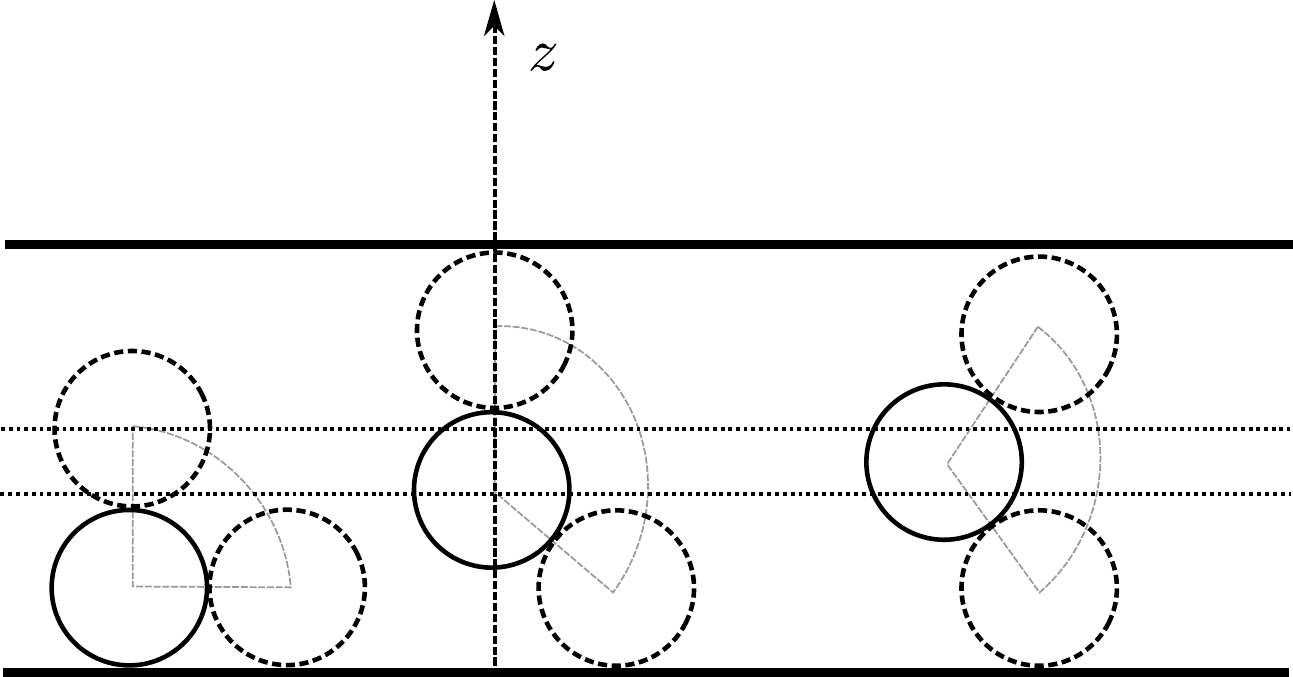} 
\end{center}
\caption{Diagram of the possible solid angles for three different
  heights of the tagged particle (represented by a solid line circle)
  for $2\sigma \leq h  \leq 3 \sigma $. The dashed line circles
  represent particles colliding with the tagged particle with the
  maximum and minimum value of $\theta$. If the center of the tagged
  particle is in the intermediate region between the two dashed lines
  ($h-3\sigma/2 \leq h \leq 3\sigma/2$), the two walls restrict the
  possible angles (right side). If the center of the tagged particle
  is between the dashed line and the closest wall, only that wall
  restricts (center and left side).  } 
\label{fig:4}
\end{figure}

Here, depending on the height of the tagged particle, one or the two
walls restrict the orientation of the possible collisions. If the
tagged particle is closed to one wall, concretely, if the distance
between the center of the particle to one of the walls is smaller than
$h-3\sigma/2$, only that wall restricts the orientation of the
possible collisions. There is also an intermediate region ($h-3 \sigma
/ 2 < z < 3\sigma /2$) where the two walls contribute. In
Fig.~\ref{fig:4}, a diagram of the possible solid angles for three
different heights of the tagged particle is represented. Then, the
allowed solid angles can be parametrized as follows  
\begin{align}\label{eq:intpar} \Omega_h( z) \equiv \left\lbrace 
\begin{aligned}
    \begin{split}
           \{ (\theta, \phi)|\;\theta\in (0, \pi/2+b_1(z)), \phi\in
           (0,2\pi) \} ,\; & \hbox{ for } \sigma/2<z<h-3\sigma/2,
           \\ 
            \{ (\theta, \phi)|\;\theta \in (\pi/2-b_2(z, h), \pi/2+b_1(z)), \phi\in
            (0,2\pi)\} ,\;  &\hbox{ for } 
            h-3\sigma/2<z<3\sigma/2 ,\\ 
            \{ (\theta, \phi)|\;\theta \in (\pi/2-b_2(z,h), \pi ),
            \phi\in(0, 2\pi)  \} ,\; & \hbox{ for }  3\sigma/2<z<h-\sigma/2. \\
            \end{split}
    \end{aligned}
    \right.
\end{align}

%%% wide systems %% 

\begin{itemize}
\item \ul{$ h \geq 3 \sigma$}
\end{itemize}
\begin{figure}[t!]
\centering 
\includegraphics[scale=0.5]{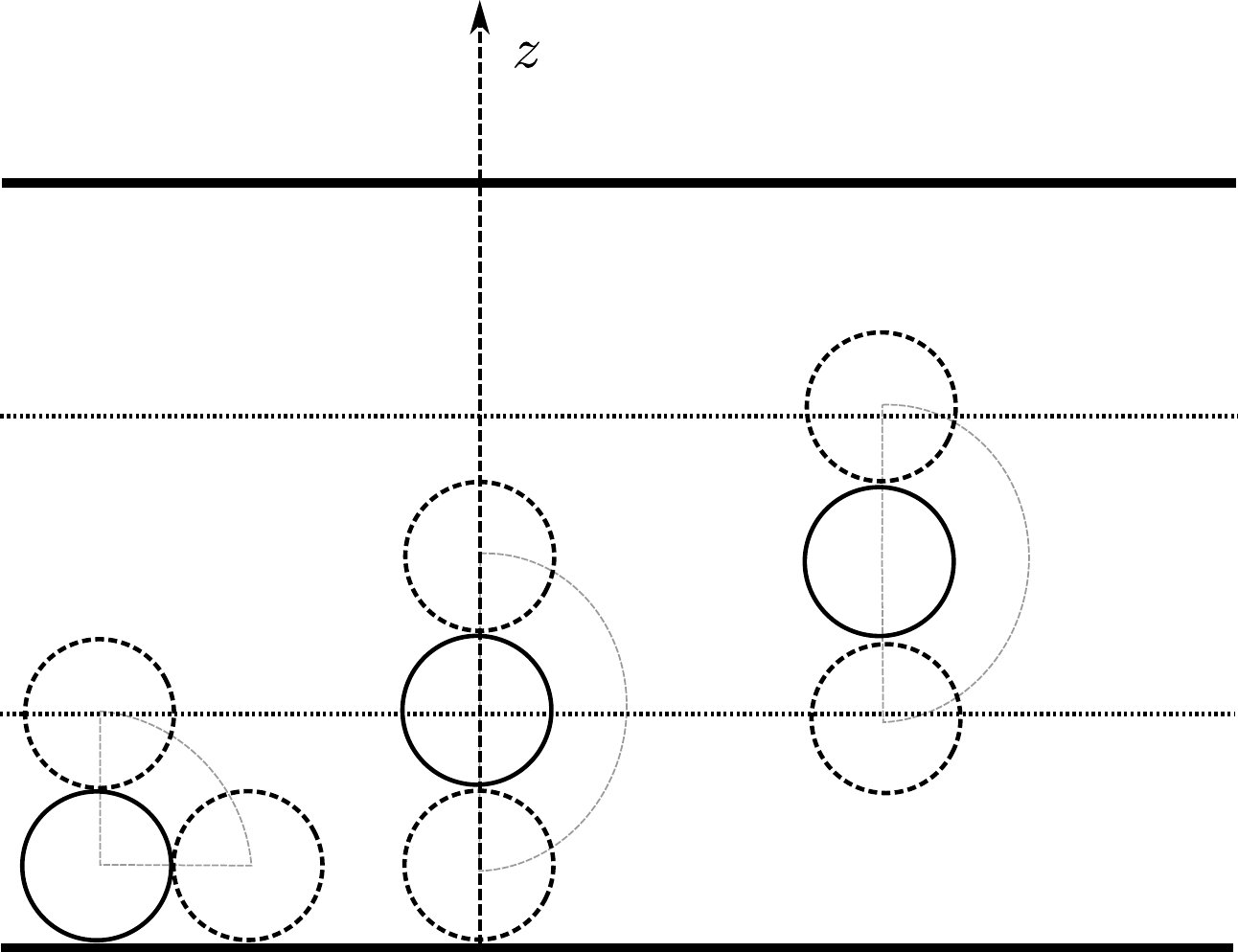} 
\caption{Diagram of the possible solid angles for three different
  heights of the tagged particle (represented by a solid line circles)
  for $ h  \geq 3 \sigma $. The dashed line circles represent
  particles colliding with the tagged particle with the maximum and
  minimum value of $\theta$. In the intermediate region between the
  two dashed lines ($3\sigma/2 \leq h \leq h- 3\sigma/2$), there is no
  restriction on the possible angles (right side). If the center of
  the tagged particle is between the dashed line and the closest wall,
  only that walls restricts (center and left side). } 
\label{fig:5}
\end{figure}

Finally, in this case there is always an intermediate region ($3\sigma/2 <z < h -3\sigma /2$) where there is no restriction on the angles of collision and that can be considered to be the ``bulk'' of the system. Outside this region, the closest wall restricts the possible angles of collisions. In Fig.~\ref{fig:5}, a diagram of the possible solid angles for three different heights of the tagged particle is represented. The allowed solid angles can be parametrized as follows 
\begin{align}\label{ecu9} \Omega_h( z) \equiv \left\lbrace 
\begin{aligned}
    \begin{split}
           \{ (\theta, \phi)|\;\theta\in (0, \pi/2+b_1(z)), \phi\in
           (0,2\pi) \} ,\; & \hbox{ for }\sigma/2<z<3\sigma/2,
           \\ 
            \{ (\theta, \phi)|\;\theta \in (0, \pi), \phi\in
            (0,2\pi)\} ,\;  &\hbox{ for } 3\sigma/2<z<h-3\sigma/2 ,\\ 
            \{ (\theta, \phi)|\;\theta \in (\pi/2-b_2(z,h), \pi ),
            \phi\in(0, 2\pi)  \} ,\; & \hbox{ for } h-3\sigma/2<z<h-\sigma/2. \\
            \end{split}
    \end{aligned}
    \right.
\end{align}

%The kinetic equation~\eqref{eq:bteq2d} has to be solved with proper boundary conditions in the horizontal direction. Assuming that $ \bm{r}_\parallel \in  A $ and vertical elastic walls at $ \bm{r}_\parallel \in   \partial A $, the distribution function fulfils
%\begin{align}\label{boundary_cond}
%\overline{f}(\bm{r}_\parallel,\bm{v},t)=\overline{f}(\bm{r}_\parallel,b_{N}\bm{v},t),\; \forall \bm{r}_\parallel\in \partial A,  \forall \bm{v},\forall t,
%\end{align} 
%where $b_N$ is the operator that inverts the sign of the horizontal components of velocity,
%\begin{align}
%b_N \bm{v} \equiv  \bm{v}-2\left[\bm{N}(\bm{r}_\parallel)\cdot\bm{v}\right] \bm{N}(\bm{r}_\parallel),
%\end{align}
%with $\bm{N}(\bm{r}_\parallel)  $ the normal vector to the boundary (see Fig.~\ref{fig:6}).
%
%\begin{figure}[h!]
%\includegraphics[scale=0.6]{figures/patatoide.pdf}
%\caption{ The top view of the system (left) where the boundary of the system is $\partial  A $ and its normal vector at $\bm{r}_\parallel$ is $\bm{N}(\bm{r}_\parallel)$. Lateral view of the system (right) with the particle colliding with the vertical wall located at $\bm{r}_\parallel$ with normal vector $\bm{N}(\bm{r}_\parallel)$. The precollisional and poscollisional with the wall velocities, $\bm{v}$ and $b_N\bm{v}$, are also represented. }
%\label{fig:6}
%\end{figure}

The kinetic equation admits as a solution the equilibrium distribution function
\begin{align} \label{eq:eq_df}
 \overline{f}_{\text{eq}} (\bm{v} ) \equiv   n \chi_\text{M}(\bm{v}),
\end{align}
where
\begin{equation}
\chi_\text{M} (\bm{v})\equiv \left(\frac{m}{2 \pi k_B T }\right)^{3/2}e^{-\frac{mv^2}{2 k_B T }},
\end{equation}
is the Maxwellian distribution with temperature $T$, and  $n \equiv
N/A(h-\sigma)$, the effective 3d density. $k_B$ is the
Boltzmann constant.  In effect,
$\mathcal{J}_h[\overline{f}_{\text{eq}} ,\overline{f}_{\text{eq}} ]$
trivially vanishes, due to the fact that $(b_{{\widehat{\bm\sigma}}}
-  1 ) \overline{f}_{\text{eq}}  (\bm{v}_1 ) \overline{f}_{\text{eq}}
(\bm{v} ) = 0 $. %Let also note that this solution trivially satisfies
%the boundary conditions.  

\subsection{Some properties of the collision operator}

Let us summarize some properties of the collision operator that are
useful in this context and that will be extended to the context of the
Boltzmann-Lorentz equation latter on.   

In Appendix \ref{appendixa} it is proved that, as in the Boltzmann
equation for bulk systems, the theta function in the collision operator
$\Theta(-\bm{g} \cdot\widehat{\boldsymbol{\sigma}})$ can be
replaced by $\Theta(\bm{g}\cdot\widehat{\boldsymbol{\sigma}})$, so
that the collisional operator can be written in the more symmetric
form 
\begin{equation} \label{eq:col_op2}
\begin{aligned}
\mathcal{J}_h[\overline{f}|\overline{f}]=&\frac{\sigma^{2}}{2(h-\sigma)}  \int_{\sigma / 2}^{h-\sigma / 2} \dd z  \int   \dd \bm{v}_1  \underset{\Omega_h(z) }{ \int} \dd \un|\bm{g} \vdot \un|(b_{{\widehat{\bm\sigma}}}  -  1 ) \overline{f}\left(\bm{r}_{\|}, \bm{v}_1, t\right) \overline{f}\left(\bm{r}_{\|}, \bm{v}, t\right).
\end{aligned}
\end{equation}
In appendix~\ref{appendixa} it is also proved that 
\begin{equation}\label{eq:bprop11}
\begin{aligned}
\int \dd \boldsymbol{v} \psi(\boldsymbol{v}) \mathcal{J}_h[\overline{f}|\overline{f}]
=&  - \frac{ \sigma^{2}}{8(h-\sigma)}\int_{\sigma/2}^{h-\sigma/2} \dd
z \int \dd \boldsymbol{v} \int \dd \boldsymbol{v}_1 \int_{\Omega_h(z)}
\dd \un |\boldsymbol{g} \cdot \widehat{\boldsymbol{\sigma}}|
(b_{{\widehat{\bm\sigma}}}  -  1 ) \overline{f}\left(\bm{r}_{\|},
  \bm{v}_1, t\right) \overline{f}\left(\bm{r}_{\|}, \bm{v}, t\right)
\\ 
& \hspace{150pt} (b_{{\widehat{\bm\sigma}}}  -  1
)\left[\psi\left(\boldsymbol{v}\right)+\psi(\boldsymbol{v}_1)\right], 
\end{aligned}
\end{equation}
where $\psi(\bm{v})$ is an  arbitrary function of the
velocity. Eq.~\eqref{eq:bprop11} can alternatively be written in the
form 
\begin{equation}\label{eq:bprop12}
\begin{aligned}
\int \dd \boldsymbol{v} \psi(\boldsymbol{v})
\mathcal{J}_h[\overline{f}|\overline{f}]= &  \frac{
  \sigma^{2}}{4(h-\sigma)}\int_{\sigma/2}^{h-\sigma/2} \dd z \int \dd
\boldsymbol{v} \int \dd \boldsymbol{v}_1 \int_{\Omega_h(z)} \dd \un
|\boldsymbol{g} \cdot \widehat{\boldsymbol{\sigma}}|
\overline{f}\left(\bm{r}_{\|}, \bm{v}_1, t\right)
\overline{f}\left(\bm{r}_{\|}, \bm{v}, t\right) \\ 
& \hspace{200pt} (b_{{\widehat{\bm\sigma}}}  -  1 )
\left[\psi\left(\boldsymbol{v}\right)+\psi(\boldsymbol{v}_1)\right]. 
\end{aligned}
\end{equation}
The quantity $\int \dd \boldsymbol{v} \psi(\boldsymbol{v})
\mathcal{J}_h[\overline{f}|\overline{f}]$ can be interpreted as the
collisional contribution to the time variation of the local quantity
$\int \dd \boldsymbol{v} \psi(\boldsymbol{v})
\overline{f}\left(\bm{r}_{\|}, \bm{v}, t\right)$. So,
Eq.~\eqref{eq:bprop12} is transparent in the sense that it indicates
that this contribution is, basically, the mean value of the increment
$(b_{{\widehat{\bm\sigma}}}  -  1 )
\left[\psi\left(\boldsymbol{v}\right)+\psi(\boldsymbol{v}_1)\right]
$. Of course,  $\int \dd \boldsymbol{v} \psi(\boldsymbol{v})
\mathcal{J}_h[\overline{f}|\overline{f}]=0$ if $\psi(\bm{v})$ is a
collisional invariant, i.e., $\psi(\bm{v}) = 1$, $v_x,v_y,v_z$ or
$v^2$.   

For $h\geq 2 \sigma$, the collision operator can be decomposed into a
``bulk'' part, where there is no limitation in the solid angles, plus
a contribution corresponding to a system with $h=2\sigma$. A similar
property was proved in \cite{mpgm23} in the context of confined
granular gases. In Appendix \ref{apendixc} it is proved that 
\begin{align}\label{eq:dec_ct}
\mathcal{J}_h[\overline{f}|\overline{f}] = \frac{\ell-1}{\ell} \mathcal{J}^{(b)}[\overline{f}|\overline{f}] + \ell^{-1}\mathcal{J}_{2\sigma} [\overline{f}|\overline{f}],
\end{align}
where 
\begin{align}\label{eq:ad_al}
\ell \equiv \frac{h-\sigma}{\sigma},
\end{align}
is the dimensionless accessible height,
\begin{equation}
 \mathcal{J}^{(b)}[\overline{f}|\overline{f}] \equiv  \frac{ \sigma^{2}}{2}  \int   \dd \bm{v}_1  \underset{\Omega }{ \int} \dd \un|\bm{g} \vdot \un| (b_{\un} -1) \overline{f}(\bm{r}_\parallel,\bm{v}_1,t)\overline{f}(\bm{r}_\parallel,\bm{v},t),
\label{ecu16}
\end{equation}
is the bulk collisional contribution because there is no limitation in the solid angle domain, $\Omega\equiv \{\theta\in(0,\pi),\phi\in(0,2\pi)\}$, and 
\begin{equation}
 \mathcal{J}_{2\sigma}[\overline{f}|\overline{f}] \equiv  \frac{ \sigma}{2}  \int^{3\sigma/2}_{\sigma/2} \dd z   \int   \dd \bm{v}_1  \underset{\Omega_{2\sigma}(z) }{ \int} d \un|\bm{g} \vdot \un| (b_{\un} -1) \overline{f}(\bm{r}_\parallel,\bm{v}_1,t)\overline{f}(\bm{r}_\parallel,\bm{v},t),
\label{ecu17}
\end{equation}
is the collisional contribution of a confined system of height $2\sigma$. 
This property is very useful because let us study the dynamics of a
system with $h>2\sigma$ in terms of the dynamics of a system of height
$2\sigma$, plus a bulk contribution of a system of height
$h-2\sigma$. For $h\sim 2 \sigma 
$, the first term in Eq.~\eqref{eq:dec_ct} is negligible with respect
the second one and the collisional contribution is, essentially, the
one of a $2 \sigma$-height system. For large height system, the
collisional contribution is, basically, the one of a pure bulk
system. Interestingly, in the intermediate region the collisional
contribution is the conveniently weighted sum of the two contributions
given by Eq.~\eqref{eq:dec_ct}. 

\subsection{$H$-Theorem}
In this section an $H$-theorem will be proved. In order to do that,
the same steps than in the Boltzmann equation for bulk systems are
followed
\cite{resibois1977classical,soto2016kinetic,dorfman2021contemporary}. First,
let us define the functional 
\begin{align}\label{eq:h_func}
\mathcal{H}[\overline{f}] \equiv \int_A \dd \bm{r}_\parallel \int \dd \bm{v} \overline{f}(\bm{r}_\parallel, \bm{v},t)\ln \overline{f}(\bm{r}_\parallel, \bm{v},t).
\end{align} 
Taking the derivative with respect time of Eq.~\eqref{eq:h_func}
 \begin{align}
\begin{aligned}\label{eq:d2}
\dv{\mathcal{H}}{t} &= \int_A \dd \bm{r}_\parallel \int \dd \bm{v} \left(\pdv{\overline{f}(\bm{r}_\parallel, \bm{v},t)}{t}\ln \overline{f}(\bm{r}_\parallel, \bm{v},t)+\pdv{\overline{f}(\bm{r}_\parallel, \bm{v},t)}{t}\right)\\
&=  \int_A \dd \bm{r}_\parallel \int \dd \bm{v} \ln \overline{f}(\bm{r}_\parallel, \bm{v},t)  \left(-\bm{v}_\parallel \cdot \nabla_{\parallel} \overline{f}(\bm{r}_\parallel, \bm{v},t) + \mathcal{J}_h[\overline{f}|\overline{f}]\right),
\end{aligned}
\end{align}
where we have used the conservation of the number of particles and the
kinetic equation, Eq.~\eqref{eq:bteq2d}. The first term of
Eq.~\eqref{eq:d2} can be rewritten as 
 \begin{align}
\begin{aligned}\label{eq:d3}
- \int_A \dd \bm{r}_\parallel \int \dd \bm{v} \ln \overline{f}(\bm{r}_\parallel, \bm{v},t)  \bm{v}_\parallel \cdot \nabla_{\parallel} \overline{f}(\bm{r}_\parallel, \bm{v},t) =&\int_A \dd \bm{r}_\parallel \int \dd \bm{v} \nabla_{\parallel} \cdot \left( \bm{v}_\parallel \overline{f}(\bm{r}_\parallel, \bm{v},t) \ln \overline{f}(\bm{r}_\parallel, \bm{v},t) -\bm{v}_\parallel \overline{f}(\bm{r}_\parallel, \bm{v},t) \right)\\
 =&\int_{\partial A} \dd \bm{S} \cdot \int \dd \bm{v}  \bm{v}_\parallel \overline{f} (\bm{r}_\parallel, \bm{v},t) \left( \ln \overline{f}(\bm{r}_\parallel, \bm{v},t) -1 \right)=0,
\end{aligned}
\end{align}
because it is assumed that there is no flux through the lateral
boundary. Hence, only collisions contribute to the time derivative of
$\mathcal{H}$ 
 \begin{align}
\begin{aligned}\label{eq:d4}
\dv{\mathcal{H}}{t} &=\int_A \dd \bm{r}_\parallel \int \dd \bm{v} \ln  \overline{f}   (\bm{r}_\parallel,\bm{v},t) \mathcal{J}_h[\overline{f}|\overline{f}].
\end{aligned}
\end{align}
Finally, by using Eq.~\eqref{eq:bprop11}, we have
 \begin{align}
\begin{aligned}\label{eq:d5}
\dv{\mathcal{H}}{t} 
&=-\frac{\sigma^2}{8(h-\sigma)}\int_A\dd \bm{r}_\parallel \int_{\sigma/2}^{h-\sigma/2} \dd z \int \dd \boldsymbol{v} \int \dd \boldsymbol{v}_1 \int_{\Omega_h(z)} \dd \un |\boldsymbol{g} \cdot \widehat{\boldsymbol{\sigma}}| \ln \frac{\overline{f}(\bm{r}_\parallel,\bm{v}^\prime,t)\overline{f}(\bm{r}_\parallel,\bm{v}^\prime_1,t)}{\overline{f}(\bm{r}_\parallel,\bm{v},t)\overline{f}(\bm{r}_\parallel,\bm{v}_1,t)}\\
&  \hspace{270pt} (b_{{\widehat{\bm\sigma}}}  -  1 )  \overline{f}(\bm{r}_\parallel,\bm{v}_1,t)\overline{f}(\bm{r}_\parallel,\bm{v},t) \\
 &\leq 0,
\end{aligned}
\end{align}
because $(b-a)\ln b/a \geq 0$, for $a$, $b>0$. I fact, it only vanishes if $a=b$. So, we can conclude that
\begin{align}\label{eq:d6}
\begin{aligned}
\dv{\mathcal{H}}{t} &=0 \iff  (b_{{\widehat{\bm\sigma}}}  -  1 )  \overline{f}(\bm{r}_\parallel,\bm{v}_1,t)\overline{f}(\bm{r}_\parallel,\bm{v},t)  =0, \\
& \hspace{100pt}\; \forall \bm{r}_\parallel,\forall \bm{v},\forall \bm{v}_1,\forall t, \forall \un \in \Omega_h(z) \text{ with } z\in[\sigma/2,h-\sigma/2]. 
\end{aligned}
\end{align}  
This condition implies that, if the time derivative of $\mathcal{H}$
vanishes, $\ln \overline{f}(\bm{r}_\parallel,\bm{v},t)$ is a linear
combination of collisional integrals, or equivalently, a local
equilibrium distribution function \cite{resibois1977classical}. This
condition plus the fact that functional $\mathcal{H}$ is lower bounded
implies that $\overline{f}\to \overline{f}_{\text{eq}}$ for $t\gg$, as
in the Boltzmann equation for bulk systems. Note that the result was
expected to hold as a consequence of the $H$-theorem derived in
\cite{bmg16} for the complete distribution function, $f(\bm{r}, \bm{v},
t)$.

%%%%%%%%%%%%%%%%% Boltamznn-Lorentz and the diffusion eq. %%%%%%%%%%%%
\section{Boltzmann--Lorentz equation and self-diffusion}
\label{sec:bl}
In the following, we will study the self-diffusion process. The
objective is to study the dynamics of some particles of the gas that
are tagged by, for example, coloring them. By imposing the condition that the system
is, as a whole, in equilibrium, the velocity distribution of the
tagged particles, $f_\text{l}(\bm{r}_\parallel,\bm{v},t)$ (the bar in
the distribution function has
been eliminated for simplicity), verifies
the following kinetic equation 
\begin{equation}
\left(\frac{\partial}{\partial t} 
+ \bm{v}_\parallel\cdot \nabla_\parallel \right) f_\text{l}(\bm{r}_\parallel,\bm{v},t) 
= \Lambda_h(\bm{v}) f_\text{l}(\bm{r}_\parallel,\bm{v},t) .
\label{ecu5}
\end{equation}
Here the collision operator, $\Lambda_h(\bm{v})$, is linear and can be written as 
\begin{equation}
\Lambda_h(\bm{v}) f_{\text{l}}(\bm{r}_\parallel,\bm{v},t) =\frac{n
  \sigma^{2}}{2(h-\sigma)}  \int_{\sigma / 2}^{h-\sigma / 2} \dd z
\int   \dd \bm{v}_1  \underset{\Omega_h(z) }{ \int} \dd \un|\bm{g}
\vdot \un|(b_{{\widehat{\bm\sigma}}}  -  1 ) \chi_\text{M}
(\bm{v}_1)f_\text{l}(\bm{r}_\parallel,\bm{v},t) , 
\label{ecu6}
\end{equation}
where the parametrization of $\Omega_h(z)$ is given in the previous
section for the different values of $h$. Eq.~\eqref{ecu5} is the
generalization of the so called Boltzmann-Lorentz (B-L) equation
\cite{resibois1977classical,soto2016kinetic,dorfman2021contemporary},
to our model for confined hard spheres. The main difference with
respect to the B-L equation for bulk systems is the concrete form of
the linearized collision operator, that takes into account the effect
of the confinement.
 
The B-L collision operator verifies similar properties that the
non-linear collision operator. The corresponding property to the one
given by Eq. (\ref{eq:bprop12}) is 
 \begin{equation} \label{eq:blproperty1}
\begin{aligned}
\int \dd \boldsymbol{v} \psi(\boldsymbol{v}) \Lambda_h(\bm{v}) f_\text{l} (\bm{r}_\parallel,\boldsymbol{v},t)= &  \frac{n \sigma^{2}}{2(h-\sigma)}\int_{\sigma/2}^{h-\sigma/2} \dd z \int \dd \boldsymbol{v} \int \dd \boldsymbol{v}_1 \int_{\Omega_h(z)} \dd \un |\boldsymbol{g} \cdot \widehat{\boldsymbol{\sigma}}| \chi_\text{M} \left( \boldsymbol{v}_1\right) f_\text{l} (\bm{r}_\parallel,\boldsymbol{v},t) \\
& \times\left[\psi\left(\boldsymbol{v}^{\prime}\right)-\psi(\boldsymbol{v})\right] ,
\end{aligned}
\end{equation}
so that $\int \dd \boldsymbol{v}  \Lambda_h(\bm{v})
f_\text{l}(\bm{r}_\parallel,\boldsymbol{v},t)=0$. For $h\geq 2\sigma$,
as in Eq. (\ref{eq:dec_ct}) for the non-linear operator, $\Lambda_h$
can be decomposed as 
\begin{align} \label{eq:bloperator_divided}
\Lambda_h(\bm{v}) \equiv  \frac{\ell-1}{\ell}\Lambda^{(b)}(\bm{v}) 
+\ell^{-1}\Lambda_{2\sigma}(\bm{v}) ,
\end{align}
where
\begin{equation}
\Lambda^{(b)}(\bm{v}) f_{\text{l}}(\bm{r}_\parallel,\bm{v},t) =\frac{n \sigma^{2}}{2}    \int   \dd \bm{v}_1  \underset{\Omega  }{ \int} \dd \un|\bm{g} \vdot \un|(b_{{\widehat{\bm\sigma}}}  -  1 ) \chi_\text{M}  (\bm{v}_1)f_\text{l}(\bm{r}_\parallel,\bm{v},t) ,
\label{b-l_bulk}
\end{equation}
is the B--L collision operator without confinement and $\Lambda_{2\sigma}$ the B--L collision operator of a system of height $2\sigma$. 

The objective now is to obtain an evolution equation for the averaged
density of tagged particles 
\begin{align}\label{eq:dens_of_labelled_particles}
\rho (\bm{r}_\parallel,t) \equiv \int \dd \bm{v} f_\text{l}(\bm{r}_\parallel,\bm{v},t).
\end{align}
By integrating in the velocity in the B-L equation, Eq.~\eqref{ecu5},
the following equation is obtained  
\begin{align}\label{eq:dens_cons_eq}
\frac{\partial}{\partial t}\rho (\bm{r}_\parallel,t) +
  \nabla_{\parallel} \cdot \bm{J}( \bm{r}_\parallel,t)  = 0 , 
\end{align}
where the flux of labelled particles in the horizontal plane is defined by
\begin{align}
\bm{J}(\bm{r}_\parallel,t) = \int \dd \bm{v} \bm{v}_{\parallel}  f_\text{l}(\bm{r}_{\parallel},\bm{v},t). 
\end{align}
Note that Eq.~\eqref{eq:dens_cons_eq} is not closed, because depends
on  the flux, $\bm{J}$.  To extract a closed equation for the density
of labelled particles, the projection operator method will be used
\cite{zwanzig1960,mori1965,zwanzig2001}.  
To proceed, let us introduce the scalar product 
\begin{align} \label{inner_product}
    \braket{k}{g} \equiv  \int \dd \bm{v}   \chi_{\text{M}}^{-1} (\bm{v}) k^*(\bm{v}) g(\bm{v}) ,
    \end{align}
defined in the space of functions $L^2_\chi(\mathbb{R}^3)$, for which
the corresponding norm is finite, i.e., $f \in L^2_\chi(\mathbb{R}^3)$
if $\braket{f}{f}<\infty$. With this scalar product, $\Lambda_h$ is
hermitian. In effect, by standard manipulations it can be shown that
\cite{resibois1977classical}  
\begin{align}\label{hermit}
\begin{aligned}
\bra{k}\Lambda_h \ket{g} = -\frac{n\sigma^2}{4(h-\sigma)} \int \dd \bm{v} \int \dd \bm{v}_1 \int_{\sigma /2}^{h-\sigma/2} \dd z \int |\bm{g}\cdot \widehat{\bm{\sigma}}| \chi_M (\bm{v}) \chi_M (\bm{v}_1) \\
\times  \left[\frac{k(\bm{v}^\prime)}{ \chi_M (\bm{v}^\prime)}- \frac{k(\bm{v})}{ \chi_M (\bm{v})}\right]^{*}\left[\frac{g(\bm{v}^\prime)}{ \chi_M (\bm{v}^\prime)}- \frac{g(\bm{v})}{ \chi_M (\bm{v})}\right] . 
 \end{aligned}
\end{align}
Moreover, Eq.~\eqref{hermit} also shows that $\bra{k}\Lambda_h \ket{k}\leq 0$, so that the eigenvalues of the operator are semi-negative, being 
the only eigenfunction associated to the null eigenvalue the Maxwellian distribution, $\chi_M (\bm{v})$. 

Let us introduce a projection operator that projects on to the subspace associated to the null eigenvalue 
\begin{align}
\mathcal{P} g(\bm{v}) = \braket{\chi_{\text{M}}}{g}\chi_{\text{M}}(\bm{v}) .
\end{align}
It is easily seen that $ \mathcal{P}$ is a projector as $\mathcal{P}^2 = \mathcal{P} $, because $\mathcal{P} \chi_{\text{M}}(\bm{v} ) = \chi_{\text{M}}(\bm{v})$ ($\chi_{\text{M}}(\bm{v})$ is normalized, therefore $\braket{\chi_M}{\chi_M}=1$). In fact, when applied to the distribution function, it leads to
 \begin{align}
\mathcal{P} f_\text{l}(\bm{r}_\parallel,\bm{v},t) = \rho(\bm{r}_\parallel, t) \chi_{\text{M}}(\bm{v}) ,
\end{align}
so that $\rho(\bm{r}_\parallel, t)$ can be interpreted as the
``component'' of $f_\text{l}$ in the direction of $\chi_{\text{M}}$
and Eq.~\eqref{eq:dens_cons_eq} as the projection of the B--L
equation. The objective is, then, to obtain a closed equation for
$\mathcal{P}f_{\text{l}}$.  Let us also introduce the orthogonal
projection through 
\begin{align}
\mathcal{Q} \equiv  \mathbb{I} - \mathcal{P},
\end{align}
where $ \mathbb{I}$ is the identity operator. It is convenient to work in Fourier space. After defining the Fourier transform as usual,
\begin{align}
f_{\text{l},\;\bm{k}}(\bm{v},t)\equiv \int \dd \bm{r}_\parallel e^{-i\bm{k}\cdot\bm{r}_\parallel } f_\text{l}(\bm{r}_\parallel,\bm{v},t),
\end{align}
the Fourier components, $f_{\text{l},\;\bm{k}}$, verify
\begin{align} \label{fourier1}
\left(\frac{\partial }{\partial t} + \mathcal{L}_{\bm{k}}  \right) f_{\text{l},\;\bm{k}}(\bm{v},t)=0,
\end{align}
with $\mathcal{L}_{\bm{k}}\equiv\left[i \bm{k}\cdot \bm{v} - \Lambda_h(\bm{v}) \right]$. Now, we express the distribution function as the sum of two functions, $f_{\text{l},\;\bm{k}}(\bm{v},t) = \mathcal{P}f_{\text{l},\;\bm{k}}(\bm{v},t)+ \mathcal{Q} f_{\text{l},\;\bm{k}}(\bm{v},t)$, and rewrite  Eq.~\eqref{fourier1} as two couple equations as
\begin{align}
\left(\frac{\partial }{\partial t} +\mathcal{P} \mathcal{L}_{\bm{k}}
  \mathcal{P} \right) \mathcal{P} f_{\text{l},\;\bm{k}}(\bm{v},t)=
  -\mathcal{P}\mathcal{L}_{\bm{k}} \mathcal{Q}
  f_{\text{l},\;\bm{k}}(\bm{v},t), \label{eq:hydrod_part_1}\\ 
\left(\frac{\partial }{\partial t} +\mathcal{Q} \mathcal{L}_{\bm{k}}
  \mathcal{Q} \right) \mathcal{Q} f_{\text{l},\;\bm{k}}(\bm{v},t)=
  -\mathcal{Q}\mathcal{L}_{\bm{k}} \mathcal{P}
  f_{\text{l},\;\bm{k}}(\bm{v},t).\label{eq:kinet_part_1} 
\end{align}
Eq. (\ref{eq:hydrod_part_1}) can be written in the simpler way
\begin{equation}\label{eq:hydrod_part_1a}
\frac{\partial}{\partial t}\rho_{\bm{k}}(t)=-i\bm{k}\cdot\int
  \dd\bm{v}\bm{v}\mathcal{Q} f_{\text{l},\;\bm{k}}(\bm{v},t), 
\end{equation}
while Eq.~\eqref{eq:kinet_part_1} can be formally solved as
\begin{align} \label{eq:sol_kinetic_part}
\mathcal{Q} f_{\text{l},\; \bm{k}} (\bm{v}, t ) = e^{-t
  \mathcal{Q}\mathcal{L}_{\bm{k}} \mathcal{Q} } 
\mathcal{Q} f_{\text{l},\; \bm{k}}(\bm{v},0) - \int_0^t \dd \tau
  e^{-\tau \mathcal{Q}\mathcal{L}_{\bm{k}} \mathcal{Q} }
  \mathcal{Q}\mathcal{L}_{\bm{k}} \mathcal{P}  f_{\text{l},\; \bm{k}}
  (\bm{v}, t-\tau ), 
\end{align} 
in terms of the initial condition, $\mathcal{Q} f_{\text{l},\;
  \bm{k}}(\bm{v},0)$, and $\mathcal{P} f_{\text{l},\;
  \bm{k}}(\bm{v},t)$. In the long-time limit, the initial condition
term vanishes and $\mathcal{Q} f_{\text{l},\;
  \bm{k}}(\bm{v},t)$ can be expressed as a functional of $\mathcal{P} f_{\text{l},\;
  \bm{k}}$. On this time scale, by inserting
Eq.~\eqref{eq:sol_kinetic_part} into~\eqref{eq:hydrod_part_1a}, a
closed evolution equation for $\rho_{\bm{k}}(t)$ is obtained
\begin{align}\label{eq:hydrod_part_2b}
\begin{aligned}
\frac{\partial}{\partial t}\rho_{\bm{k}}(t)= -k^2 \int
\dd\bm{v}\int_0^t\dd\tau\bm{\hat{k}} \cdot \bm{v} 
e^{-\tau \mathcal{Q}\mathcal{L}_{\bm{k}}
    \mathcal{Q} }  \mathcal{Q} \bm{\hat{k}} \cdot
  \bm{v} \chi_M(\bm{v})\rho_{\bm{k}}(t-\tau),  
\end{aligned}
\end{align}
where $\bm{\hat{k}}$ is a unit vector in the direction of
$\bm{k}$. Finally, by taking the Navier-Stokes hydrodynamic limit in
Eq. (\ref{eq:hydrod_part_2b}), i.e., for long times and to second
order in $k$ with $k^2t$ 
finite \cite{zwanzig1960} and taking 
into account that $\Lambda_h$ is invariant under rotations in the
($v_x,v_y$)--plane, the following equation for $\rho_{\bm{k}}(t)$ is
obtained
\begin{align} \label{eq:dif_eq2}
\begin{aligned}
\frac{\partial }{\partial t}  \rho_{\bm{k}}(t) =&- D k^2  \rho_{\bm{k}}(t),
\end{aligned}
\end{align}
where 
\begin{align} \label{eq:dif_coef}
D = \frac{1}{2}\int_0^\infty \dd t \int \dd \bm{v} \bm{v}_\parallel\cdot
  e^{t \Lambda_h (\bm{v})} \bm{v}_\parallel \chi_M(\bm{v}). 
\end{align}
Eq.~\eqref{eq:dif_eq2} is a diffusion  equation for the 2d density in
Fourier space with a diffusion coefficient, $D$,  given by
Eq.~\eqref{eq:dif_coef}. Eq.~\eqref{eq:dif_coef} represents the 
Green-Kubo formula for $D$ \cite{green1954,kubo1957} where
$\frac{1}{2}\int \dd \bm{v} \bm{v}_\parallel\cdot
  e^{t \Lambda_h (\bm{v})} \bm{v}_\parallel \chi_M(\bm{v})$ can
be identified as the Boltzmann approximation of the correlation
function of the velocity that decays 
in the kinetic scale, i.e., with the strictly negative 
eigenvalues of $\Lambda_h$. Let us remark that the actual velocity
autocorrelation function is expected to decay on a much slower time scale~\cite{dorfman2021contemporary}. Nevertheless, this occurs
in a time scale that goes beyond the one we are interested in (the one
of the simulations of the following section). By performing the integral in
Eq.~\eqref{eq:dif_coef}, the diffusion coefficient can be rewritten in
the form  
  \begin{align} \label{eq:dif_coef4}
\begin{aligned}
 D = &- \frac{1}{2}\int \dd \bm{v }  \bm{v}_\parallel \cdot \left[\Lambda_h^{-1}(\bm{v})   \bm{v}_\parallel   \chi_\text{M} (\bm{v}) \right].
\end{aligned}
\end{align}
Note that the expressions for the transport coefficient given by
Eqs.\eqref{eq:dif_coef} and~\eqref{eq:dif_coef4} have exactly the same
form as the ones for an unconfined system: The former ones are
obtained by replacing in the last ones $\Lambda^{(b)}$ by $\Lambda_h$
\cite{resibois1977classical,dorfman2021contemporary}.

In real space,
Eq.~\eqref{eq:dif_eq2} takes the form 
\begin{align}\label{eq:dif_eq_real}
\pdv{}{t} \rho(\bm{r}_\parallel ,t)  = D \nabla_\parallel^2 \rho(\bm{r}_\parallel,t).  
\end{align}
The mean square displacement (MSD) in the horizontal direction of the
tagged particles is defined as 
\begin{align} \label{msd}
\langle r^2_{\parallel} \rangle \equiv \frac{1}{N_{\text{l}}} \int \dd \bm{r}_\parallel r_\parallel^2 \rho(\bm{r}_\parallel, t) ,
\end{align}
where $N_\text{l}$ is the total number of tagged particles. By taking
moments in Eq.~\eqref{eq:dif_eq_real}, for long times, one obtains
\begin{align}\label{msd2}
\langle r_\parallel^2\rangle \sim  4 D t.
\end{align}

The explicit evaluation of the diffusion coefficient by
Eq.~\eqref{eq:dif_coef4} is complicated because it involves the
inversion of the operator $\Lambda_h$, that is equivalent to solving a
complex integral equation. Nevertheless, for the so-called Maxwell
molecules model, in which the collision frequency between particles
is assumed to be independent of the relative velocity 
\cite{resibois1977classical,dorfman2021contemporary}, 
$\chi_M \bm{v}_\parallel$ is an eigenfunction of the B-L
operator and the calculation can be performed exactly. Inspired by
this fact, let us calculate the diffusion coefficient by performing
the following approximation  
\begin{align}\label{eigen_approx}
\Lambda_h(\bm{v}) \chi_M \bm{v}_\parallel \simeq \gamma \chi_M \bm{v}_\parallel,
\end{align} 
i.e., assuming that  $\chi_M \bm{v}_\parallel$ is an approximate
eigenfunction of $\Lambda_h$ (this kind of approximation was already
performed in the context of granular gases \cite{bmg2011}, obtaining
very accurate expressions for the transport coefficients).  Under this
approximation, the diffusion coefficient is 
\begin{align}
D = - \frac{1}{2\gamma } \int \dd \bm{v} v_\parallel^2 \chi_M .
\end{align}
The eigenvalue, $\gamma$, can be calculated approximately by taking the first velocity moment in Eq.~\eqref{eigen_approx}, obtaining
\begin{align}\label{eigenvalue1}
 \gamma = \frac{ \int \dd \bm{v }  \bm{v}_\parallel \cdot \left[\Lambda_h(\bm{v})   \bm{v}_\parallel   \chi_\text{M} (\bm{v}) \right]}{ \int \dd \bm{v }  v_\parallel^2
  \chi_\text{M} (\bm{v})} .
\end{align}
Let us note that, in this particular case, the approximation given by Eq. 
(\ref{eigen_approx}) coincides with first Sonine approximation typically used in
kinetic theory \cite{dorfman2021contemporary}. In Appendix
\ref{apendixf}, $\gamma$ is calculated 
using Eq.~\eqref{eigenvalue1}, obtaining the following expression for
$D$ 
 \begin{align} \label{eq:sel_diff}
  D = & D_0  D^*(\ell) ,
 \end{align}
 being
 \begin{align}
 D_0 = \frac{3}{8\sqrt{\pi}n\sigma^2}\left(\frac{k_B T  }{ m }\right)^{1/2},
 \end{align}
 the 3d diffusion coefficient and 
 \begin{align} \label{total_diff}
 D^*(\ell) =  \left\lbrace \begin{matrix}
\frac{8}{\ell(6-\ell^2)} & \text{ if } 0\leq \ell\leq 1, \\
\frac{8 \ell}{8 \ell-3} & \text{ if } \ell \geq 1,
\end{matrix} \right. 
\end{align}
a dimensionless function   of $\ell$ that takes into account the
effect of the confinement.

Eq.~\eqref{eq:sel_diff} is the main result
of the paper. It expresses the self-diffusion coefficient as a function of the
dimensionless height, $\ell$, without any adjustable parameter. For
large $\ell$ the diffusion coefficient is the one of a 3d system,
while for $\ell \ll 1$,  
\begin{align}
D \simeq   \frac{A}{2 \sqrt{\pi} N  \sigma  }   \left(\frac{k_B T  }{ m }\right)^{1/2}, 
\end{align}
which coincides with the diffusion coefficient of a 2d fluid. In fact, for small heights and to second order
in $\ell$, one obtains
 \begin{align}\label{eq:approx_diff_part}
D \sim  \frac{A}{2 \sqrt{\pi} N  \sigma }   \left(\frac{k_B T  }{ m }\right)^{1/2}\left( 1 + \frac{\ell^2}{6} \right),
\end{align}
which agrees with the results obtained in \cite{bgm20} for a q2d
system. Let us remark that the analysis performed in \cite{bgm20} was 
different as the density profile in the $z$-direction was taken into
account. So, to
apply the Chapman-Enskog method and derive the 
diffusion equation, a modification of the standard method was used
\cite{l06}. In contrast, in our context, the density profile in the
$z$-direction does not appear and the self-diffusion coefficient is
calculated for any  height (with the restriction $h\sim\sigma$). 
In general, for fixed 3d density, $n$, the
diffusion coefficient decays monotonically with the height. This can
be intuitively understood as, by increasing the height, the effective
horizontal diffusion is decreased. The divergence for $\ell \ll 1$ was
expected as, to keep $n$ fixed, a very large effective 2d 
density, $N/A$, is needed. At $\ell =1$ ($h=2 \sigma$), the diffusion
coefficient is continuous as function of $\ell$. Moreover $\dv{D}{\ell}$ and
$\dv{^2D}{\ell^2}$ are continuous at $\ell =1$, while $\dv{^3
  D}{\ell^3}$ is discontinuous, so that $D(\ell)$ is not analytical at
$\ell=1$.

%%%%%%%%%%%%%%%%% Simulation results %%%%%%%%%%%%
\section{Simulation results}
\label{sec:sim}

\begin{figure}[t!]
\begin{center}
\includegraphics[scale=0.4]{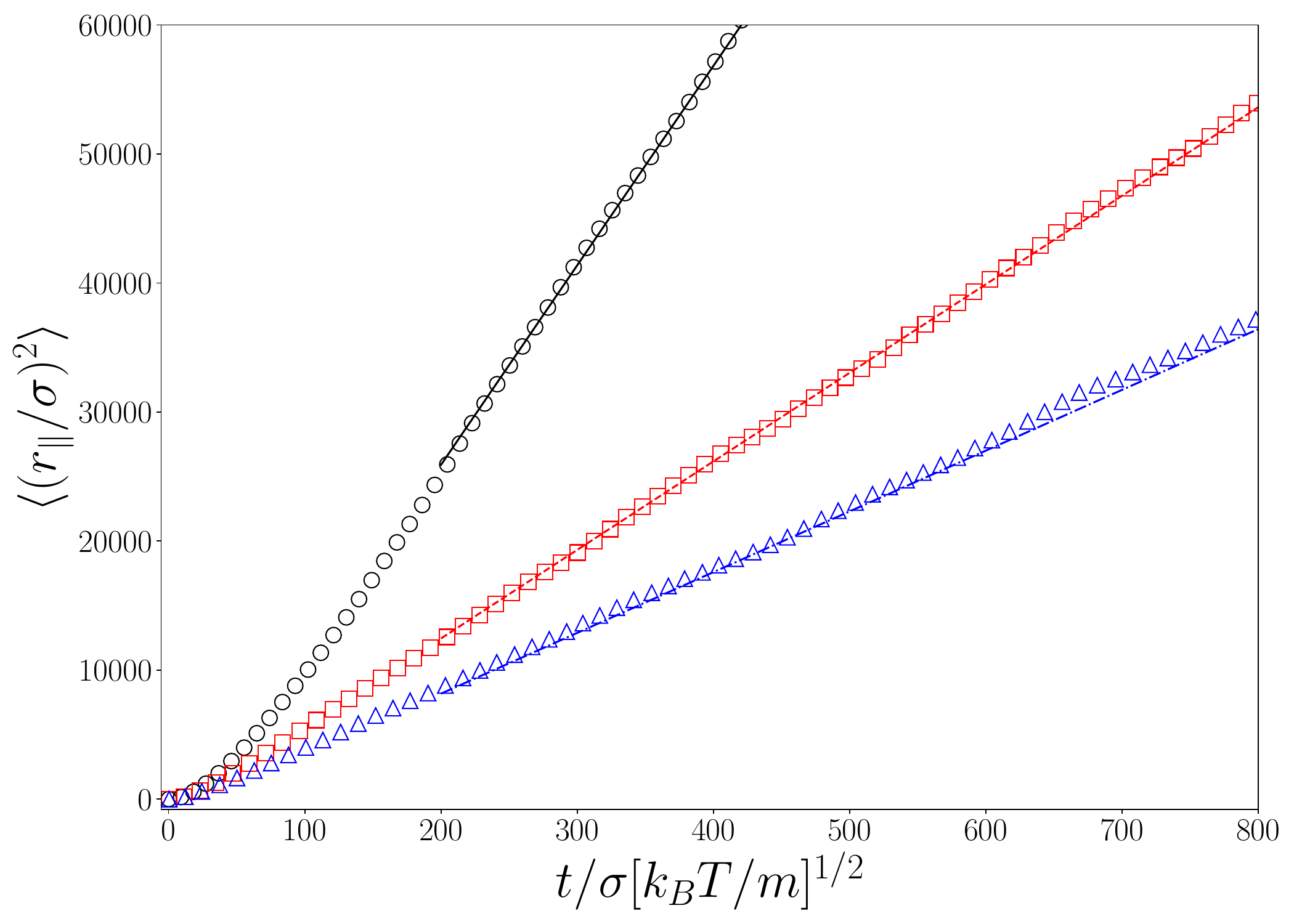}
\end{center}
\caption{ (Color online) 
$\frac{\langle r_\parallel^2 \rangle}{  \sigma^2}$ as a function of the dimensionless time, $\frac{t}{\sigma}\left(\frac{k_B T}{m }\right)^{1/2}$, for 3 values of the height (the values of the parameters are $N=500$ and $n=0.019\sigma^{-3}$).  The (black) circles, (red) squares and (blue) triangles are simulation results for $\ell = 0.4, 1 $ and $9$ respectively.  The (black) solid line, (red) dashed line and (blue) point--dashed line are the corresponding linear fitting after the initial transient. }
\label{fig:msd_different_width}
\end{figure}

\begin{figure}[t!]
\centering
\includegraphics[scale=0.4]{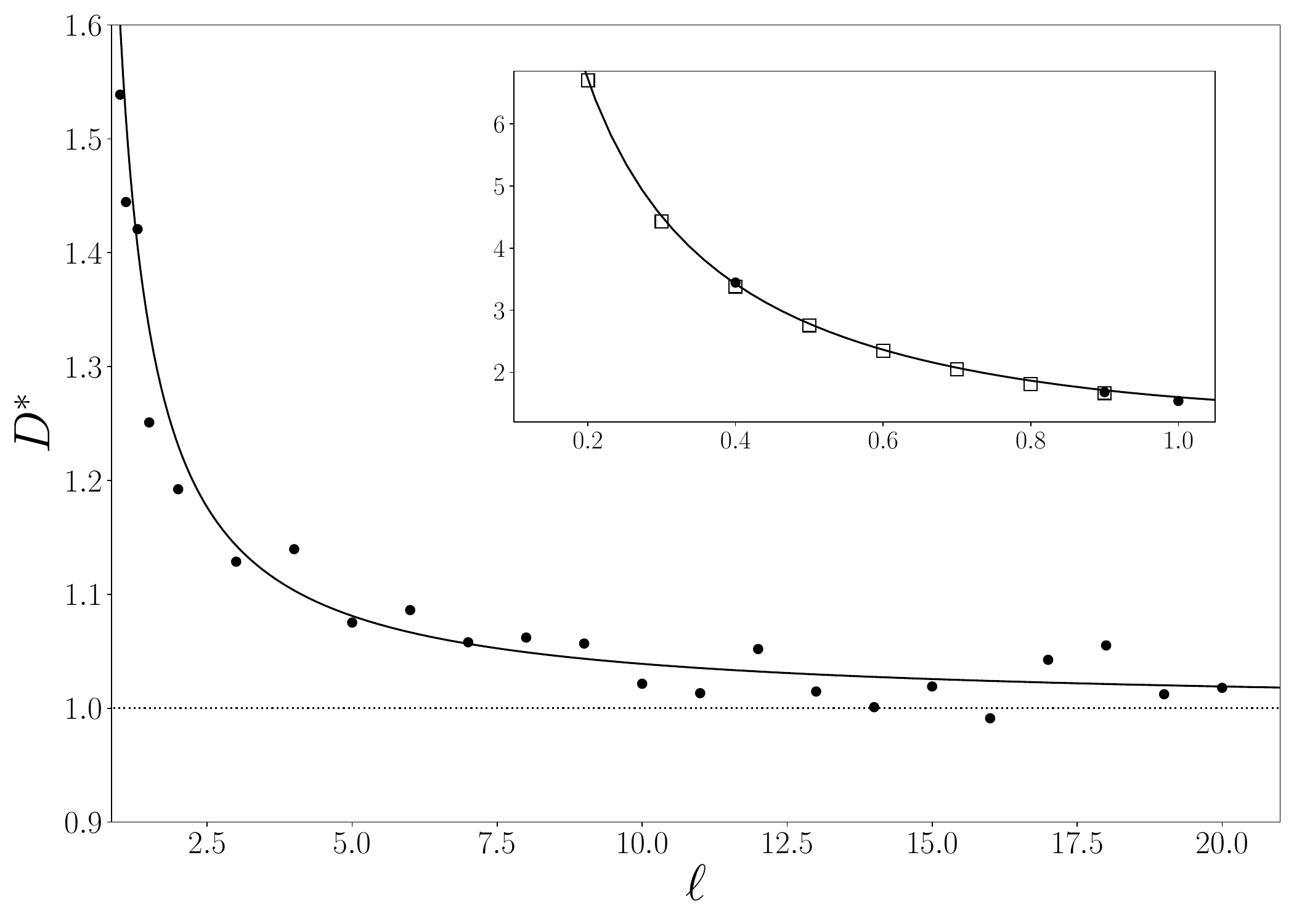
}
\caption{Dimensionless self-diffusion coefficient,
  $D^*$, as a 
  function of $\ell$. The filled circles are the MD simulation results, the solid
  line is the theoretical prediction given by Eq. (\ref{total_diff}), and the dotted
  line is the value of the dimensionless coefficient in the
  limit $\ell\to \infty$. The inset represents the behavior in the
  q2d region. The squares are the simulation results of
  Ref.~\cite{bgm20} for a q2d system.  } 
\label{fig:diff}
\end{figure}

The objective of this section is to compare the theoretical prediction
for the diffusion coefficient in the horizontal directions obtained in
the previous section with MD simulation results. Concretely, we have
measured the MSD as a function of time. As, for long times, $\langle
r_\parallel^2 \rangle \sim 4 D t$ (see Eq.~\eqref{msd2} of the
previous section), the diffusion coefficient, $D$, can be easily
extracted by linear fitting. 

We have performed MD simulations of $N$ elastic hard spheres of mass $m$ and
diameter $\sigma$, confined between two
reflecting parallel square plates of area $A$ separated a distance $h$. $m$ and
$\sigma$ are 
taken to be the units of mass and length, respectively. Periodic
boundary conditions have been 
applied in the $x$ and $y$ directions, and the sequence of collisions is generated by
the usual event driven algorithm \cite{allen2017}. Initially, all the
particles were uniformly distributed with a Maxwellian velocity
distribution of temperature $T$, that is taken to be unity. In all the
simulations, we have taken 
$N=500$ and $n=0.019 \sigma^{-3}$, so that the system can be
considered to be dilute. We have varied the height of the system from
ultraconfined 
conditions, $\ell=0.4$, to large systems, $\ell \sim 20 $, that
can be considered 3d (in the sense that boundary effects
are not expected to be relevant). The results have been averaged over
5 trajectories.   

In Fig.~\ref{fig:msd_different_width}, $\frac{\langle r_\parallel^2
  \rangle}{  \sigma^2}$ is plotted as a function of the dimensionless
time, $\frac{t}{\sigma}\left(\frac{k_B T}{m }\right)^{1/2}$, for 3
values of the height. The (black) circles, (red) squares and (blue)
triangles are the simulation results for $\ell = 0.4, 1$ and $9$,
respectively. It can be seen that, after a transient, the MSD is
approximately linear, indicating the diffusive nature of the
motion. The black (solid) line, (red) dashed line and (blue)
point--dashed line are the corresponding linear fitting, by which the
diffusion coefficient can be extracted. Similar results are obtained
for other values of $\ell$.  

In Fig.~\ref{fig:diff}, the dimensionless diffusion coefficient, $D^*$, is
plotted as a function of the dimensionless height, $\ell$. No error
bars are shown as they cannot be appreciated in the scale of the
figure. The filled circles are the MD simulation results, the solid
line is the theoretical prediction given by Eq. (\ref{total_diff}),
and the dotted line is the value 
of the dimensionless coefficient in the limit $\ell\to \infty$. The inset
represents the behavior in the q2d region. The squares are the
simulation results of Ref.~\cite{bgm20} for a q2d system. In this last
case, the simulations' parameters are $N=500$ with a fixed 2d
density, $N/A = 0.019 \sigma^{-2}$. The agreement between the theoretical
prediction and the MD simulation results is very good in the whole
range of heights. We think that this agreement is remarkable as the
theory accurately describes how the diffusion process changes from
confined system to 3d systems, but also from ultraconfined
conditions to 2d systems without any adjustable
parameter.

%%%%%%%%%%%%%%%%% Conclusions %%%%%%%%%%%%
\section{Conclusions}
\label{sec:conc}

In this paper, we have studied self-diffusion of a hard sphere system confined by two parallel
horizontal plates separated a distance of the order of the diameter of
the particles. The starting point is a closed kinetic equation for the
integrated in the vertical direction distribution function, valid in
the low-density limit. Hence, the state of the system is given by a
distribution function that depends on the horizontal spatial
coordinates, but on the 3d velocity. An $H$-theorem is proved, ensuring that
equilibrium is always reached in the long-time limit, independently of
the initial condition. From the kinetic equation, a B-L-like equation
can be written that describes the dynamics of the distribution
function of a tagged particle, in case the whole system is in
equilibrium. From it, a closed evolution equation for the 2d density
of tagged particles is obtained by applying the Zwanzig-Mori
projection formalism. A normal 2d diffusion equation is obtained with
a self-diffusion coefficient depending on the height of the
system. For small heights ($\ell\ll 1$), the 2d self-diffusion is
obtained and, to second order in $\ell$, the results agree with the
ones obtained in \cite{bgm20} for a q2d system. For large heights, the
3d self-diffusion coefficient is recovered. The self-diffusion
coefficient has been measured in MD simulations by analyzing the long
time behavior of the MSD. A very good agreement with the theoretical
prediction is obtained for the whole range of heights. It must be
stressed that there is not any adjustable parameter and that the
explicit expression of the self-diffusion coefficient given by
Eq. (\ref{eq:dif_coef4}) is obtained by a
single mathematical approximation, the one given by
Eq. (\ref{eigen_approx}).

\begin{figure}[t!]
\centering
\includegraphics[scale=0.4]{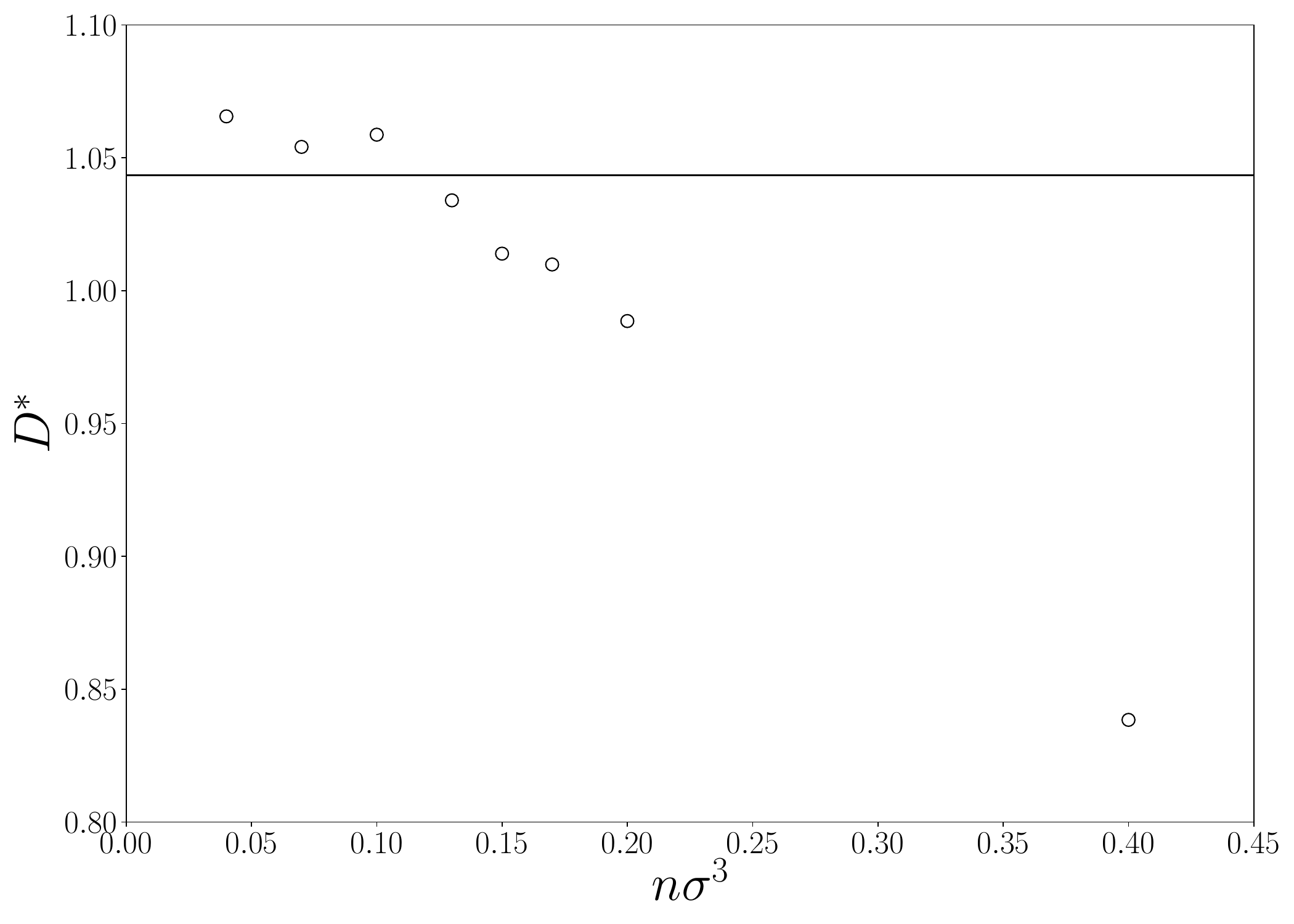}
\caption{Dimensionless diffusion coefficient, $D^*$, as a function of
  the dimensionless density, $n\sigma^3$, for $\ell =9$. The circles
  are the simulation results and the solid line is the low-density
  theoretical prediction.
} 
\label{fig:diff_dens}
\end{figure}

The theory developed here is restricted to the low-density
limit. It would be interesting to test the range of densities for
which the theory is accurate. In order to explore this point, we have
performed several MD simulations for different densities. Concretely,
we have taken a system 
with $N=1000$, fixed height at $\ell=9$, and a range of densities
varying from
$n \sigma^3 =0.05$ to $0.4$ (the results have also been averaged over 5
trajectories). In this range, we have checked that the system remains
spatially homogeneous in the bulk, i.e. that there is not a density
profile in the $z$ direction. 
%and the theory can be easily modified in the context of the Enskog
%equation by taking into account the pair correlation function at
%contact, $g_2(\sigma)$ \cite{resibois1977classical}. The new diffusion
%coefficient reads 
%\begin{align}\label{eur_dif}
%D_{\text{E}} = \frac{D}{g_2(\sigma)}.
%\end{align}
In Fig.~\ref{fig:diff_dens}, the dimensionless diffusion coefficient,
$D^*$, is plotted as a function of the dimensionless density,
$n\sigma^3$. The circles are the simulation results and the solid line
the low-density theoretical prediction given by
Eq.~\eqref{total_diff}. %The dashed line is the scaled Enskog
%theoretical prediction, $D_E$, with $g_2 (\sigma)$ given by the
%Carnahan--Starling expression \cite{Norman1969} 
%\begin{align}
%g_2(\sigma)= \frac{1-\frac{\pi}{12}n\sigma^3}{(1-\frac{\pi}{6}n\sigma^3)^3}, 
%\end{align}
It is seen that the low-density prediction agrees with the simulation
results in an unexpected wider region. For the largest density, the
simulation result is clearly smaller than the theoretical prediction,
compatible with the presence of two-particle spatial correlations that
slow down the diffusion process. Then, it would be interesting to
explore if an Enskog-like kinetic equation \cite{resibois1977classical}
accurately explains the density dependence of the self-diffusion
equation. 

In addition, there are many more open questions that arise in
  this context, beyond 
the low-density limit, e.g. recollisions induced by the walls or the
presence of long-time tails in the velocity correlation function. In
the non-confined case, the velocity 
correlation function goes as $t^{-d/2}$, being $d$ the spatial
dimensionality of the system \cite{aw70}. What happens in the q2d case?
How is the dependence on $h$? On the
other hand, in 
\cite{met06} it was found that the scaled diffusivity of a hard
sphere fluid confined by two parallel walls scaled with the excess
entropy in the same way as for non-confined fluids
\cite{d96}. Interestingly, this was the case for different wall-particle
interactions. Can this fact be understood from first principles using
kinetic theory? In the low-density limit, it would also be interesting
to explore the possibility of a hydrodynamic description in the
plane, i.e., a closed description in terms of the 2-dimensional
density, projected in the plane flow velocity and temperature. 
We think that the work presented in the paper is a
first step before these more complex questions can be addressed. 

\acknowledgments

This research was supported by grant ProyExcel-00505 funded by Junta de Andaluc\'ia, and grant PID2021-126348NB-100 funded by
MCIN/AEI/10.13039/501100011033 and ERDF ``A way of making Europe''. 

%%%%%%%%%%%%%%%%%%%%%%%%%%%%%%%%%%%%%%%%%%%%%%%%%%%%%%%%%%%%%%%%%%%%%%%%%%%%%%%

\appendix

%%%% APENDICES %%%%%%%%%%%%%%%%%%%%%%%%%%%%%%%%%%%%%%%%%%%%%%%%%%%%%%%%%%%%%%%
%%%%%%%%%%%%%%%%%%%%%%%%%%%%%%%%%%%%%%%%%%%%
\section{Some properties of the Boltzmann collision operator}
\label{appendixa}

In this Appendix, some mathematical properties concerning the
collision operator, $\mathcal{J}_h[\overline{f}|\overline{f}]$, are
proved. 

To prove that the theta function, $\Theta(-\bm{g}\cdot
\widehat{\bm{\sigma}})$, can be replaced by $\Theta(\bm{g}\cdot
\widehat{\bm{\sigma}})$, let us first prove that  
\begin{align} \label{eq:prop_int_inv}
\underset{\Omega_h(z)}{\int}\dd \un g(\un) = \underset{\Omega_h(h-z)}{\int}  \dd \un g(-\un) ,
\end{align}
 for any function of $\un$. Let us write explicitly the integral in spherical coordinates, 
 \begin{align} \label{eq:prop_inv2}
\underset{\Omega_h(z)}{\int}\dd \un g(\un) \equiv \int_{\pi/2-b_2(z,h)}^{\pi/2+b_1(z)}\dd \theta \sin \theta  \int_0^{2 \pi} \dd \phi g(\theta,\phi).
\end{align}
By performing the following change of variables
\begin{subequations}
\begin{align}
\widetilde{\theta} = \pi - \theta, \\
\widetilde{\phi} = \pi + \phi, 
\end{align}
\end{subequations}
 into Eq.~\eqref{eq:prop_inv2}, that represent the change $\un \to -\un$, it is obtained
 \begin{align} \label{eq:prop_inv3}
 \begin{aligned}
\underset{\Omega_h(z)}{\int}\dd \un g(\un) =& -
\int^{\pi/2-b_1(z)}_{\pi/2+b_2(z,h)}\dd \widetilde{\theta}  \sin (\pi
- \widetilde{\theta})  \int_\pi^{3 \pi} \dd \widetilde{\phi}
g\left(\pi - \widetilde{\theta},\pi - \widetilde{\phi}\right)\\ 
=& \int_{\pi/2-b_1(z)}^{\pi/2+b_2(z,h)}\dd \widetilde{\theta}  \sin  \widetilde{\theta}  \int_{-\pi}^{\pi} \dd \widetilde{\phi} g\left(\pi - \widetilde{\theta}, \pi+\widetilde{\phi} \right)\\
 =&\underset{\Omega_h(h-z)}{\int}  \dd \un g(-\un) ,
 \end{aligned}
\end{align}
where we have used that $b_1(z) =b_2(h-z,h)$ and $b_2(z,h)
=b_1(h-z)$. Hence, 
 \begin{align} \label{eq:prop_inv4}
\int_{\sigma/2}^{h-\sigma/2} \dd z \underset{\Omega_h(z)}{\int}\dd \un g(\un) = \int_{\sigma/2}^{h-\sigma/2} \dd z \underset{\Omega_h(z)}{\int}\dd \un g(-\un).
\end{align}
As the operator $b_{{\widehat{\bm\sigma}}} $ is invariant under the change, $ \un \to - \un $, the collisional contribution of the kinetic equation can be rewritten as 
\begin{align} \label{eq:col_op}
\begin{aligned}
\mathcal{J}_h[\overline{f}|\overline{f}]=&\frac{ \sigma^{2}}{h-\sigma}  \int_{\sigma / 2}^{h-\sigma / 2} \dd z \int   \dd \bm{v}_1  \underset{\Omega_h(z) }{ \int} \dd \un|\bm{g} \vdot \un| \Theta(-\bm{g} \vdot \un) (b_{{\widehat{\bm\sigma}}}  -  1) \overline{f}(\bm{v}_1) \overline{f}( \bm{v})\\
=&\frac{ \sigma^{2}}{h-\sigma}  \int_{\sigma / 2}^{h-\sigma / 2} \dd z \int   \dd \bm{v}_1  \underset{\Omega_h(z) }{ \int} \dd \un|\bm{g} \vdot \un| \Theta(\bm{g} \vdot \un) (b_{{\widehat{\bm\sigma}}}  -  1) \overline{f}\left( \bm{v}_1\right) \overline{f}\left( \bm{v}\right)\\
=&\frac{ \sigma^{2}}{2(h-\sigma)}  \int_{\sigma / 2}^{h-\sigma / 2} \dd z \int   \dd \bm{v}_1  \underset{\Omega_h(z) }{ \int} \dd \un|\bm{g} \vdot \un|  (b_{{\widehat{\bm\sigma}}}  -  1) \overline{f}\left( \bm{v}_1\right) \overline{f}\left( \bm{v}\right)
,
\end{aligned}
\end{align}
where the dependence of the distribution function on
$\bm{r}_\parallel$ and $t$ has been omitted.   

To prove Eqs.~\eqref{eq:bprop11} and~\eqref{eq:bprop12}, let us multiply Eq.~\eqref{eq:col_op2} by a generic function of the velocity, $\psi(\bm{v})$. After integrating in $\bm{v}$, it is obtained 
\begin{equation} \label{eq:b1}
\begin{aligned}
\int \dd \boldsymbol{v} \psi(\boldsymbol{v}) \mathcal{J}_h[\overline{f}|\overline{f}]= &  \frac{ \sigma^{2}}{2(h-\sigma)}\int_{\sigma/2}^{h-\sigma/2} \dd z \int \dd \boldsymbol{v} \int \dd \boldsymbol{v}_1 \int_{\Omega_h(z)} \dd \un |\boldsymbol{g} \cdot \widehat{\boldsymbol{\sigma}}| \psi\left(\boldsymbol{v}\right)  (b_{{\widehat{\bm\sigma}}}  -  1 )  \overline{f}(\bm{v}_1)\overline{f}(\bm{v}) .
\end{aligned}
\end{equation}
 Exchanging the label of velocities $\bm{v}\leftrightarrow\bm{v}_1$, 
\begin{equation} \label{eq:b2}
\begin{aligned}
\int \dd \boldsymbol{v} \psi(\boldsymbol{v}) \mathcal{J}_h[\overline{f}|\overline{f}]= &  \frac{ \sigma^{2}}{2(h-\sigma)}\int_{\sigma/2}^{h-\sigma/2} \dd z \int \dd \boldsymbol{v} \int \dd \boldsymbol{v}_1 \int_{\Omega_h(z)} \dd \un |\boldsymbol{g} \cdot \widehat{\boldsymbol{\sigma}}| \psi\left(\boldsymbol{v}_1\right)  (b_{{\widehat{\bm\sigma}}}  -  1 )  \overline{f}(\bm{v}_1)\overline{f}(\bm{v}) \\
=&   \frac{ \sigma^{2}}{4(h-\sigma)}\int_{\sigma/2}^{h-\sigma/2} \dd z \int \dd \boldsymbol{v} \int \dd \boldsymbol{v}_1 \int_{\Omega_h(z)} \dd \un |\boldsymbol{g} \cdot \widehat{\boldsymbol{\sigma}}|   (b_{{\widehat{\bm\sigma}}}  -  1 )  \overline{f}(\bm{v}_1)\overline{f}(\bm{v}) \\
&\left[\psi\left(\boldsymbol{v} \right) +\psi\left(\boldsymbol{v}_1\right) \right].
\end{aligned}
\end{equation}
Finally, interchanging the precollisional and poscollisional velocities, Eq.~\eqref{eq:bprop11} is obtained
 \begin{equation} \label{eq:b3}
\begin{aligned}
\int \dd \boldsymbol{v} \psi(\boldsymbol{v}) \mathcal{J}_h[\overline{f}|\overline{f}]= & 
 -  \frac{ \sigma^{2}}{4(h-\sigma)}\int_{\sigma/2}^{h-\sigma/2} \dd z \int \dd \boldsymbol{v}^\prime \int \dd \boldsymbol{v}^\prime_1 \int_{\Omega_h(z)} \dd \un |\boldsymbol{g}^\prime \cdot \widehat{\boldsymbol{\sigma}}|   (b_{{\widehat{\bm\sigma}}}  -  1 )  \overline{f}(\bm{v}_1)\overline{f}(\bm{v}) \\
&\left[\psi\left(\boldsymbol{v}^\prime \right) +\psi\left(\boldsymbol{v}^\prime_1\right) \right]\\
=& -  \frac{ \sigma^{2}}{8(h-\sigma)}\int_{\sigma/2}^{h-\sigma/2} \dd z \int \dd \boldsymbol{v} \int \dd \boldsymbol{v}_1 \int_{\Omega_h(z)} \dd \un |\boldsymbol{g}\cdot \widehat{\boldsymbol{\sigma}}|   (b_{{\widehat{\bm\sigma}}}  -  1 )  \overline{f}(\bm{v}_1)\overline{f}(\bm{v}) \\
&\left[\psi\left(\boldsymbol{v}^\prime \right) +\psi\left(\boldsymbol{v}^\prime_1\right)-\psi\left(\boldsymbol{v} \right) -\psi\left(\boldsymbol{v}_1\right) \right],
\end{aligned}
\end{equation}
where we have used that $\bm{g}'\cdot\un = - \bm{g}\cdot \un $ and
$\dd \bm{v}' \dd \bm{v}'_1=\dd \bm{v} \dd \bm{v}_1$. Another
equivalent formula can be obtained by taking into account that 
\begin{equation} \label{eq:b4}
\begin{aligned}
\int \dd \boldsymbol{v} \int \dd \boldsymbol{v}_1 \int_{\Omega_h(z)} \dd \un |\boldsymbol{g} \cdot \widehat{\boldsymbol{\sigma}}|    \overline{f}(\bm{v}^\prime_1)\overline{f}(\bm{v}^\prime) \left[\psi\left(\boldsymbol{v} \right) +\psi\left(\boldsymbol{v}_1\right) \right] =&  \int \dd \boldsymbol{v}^\prime \int \dd \boldsymbol{v}^\prime_1 \int_{\Omega_h(z)} \dd \un |\boldsymbol{g}^\prime \cdot \widehat{\boldsymbol{\sigma}}|    \overline{f}(\bm{v}_1)\overline{f}(\bm{v})\\
 &\left[\psi\left(\boldsymbol{v}^\prime \right)
   +\psi\left(\boldsymbol{v}^\prime_1\right) \right]. 
\end{aligned}
\end{equation}
By substituting Eq. (\ref{eq:b4}) into Eq.~\eqref{eq:b3},
Eq.~\eqref{eq:bprop12} is obtained, i.e.  
\begin{equation} \label{eq:b5}
\begin{aligned}
\int \dd \boldsymbol{v} \psi(\boldsymbol{v}) \mathcal{J}_h[\overline{f}|\overline{f}]= &  \frac{ \sigma^{2}}{4(h-\sigma)}\int_{\sigma/2}^{h-\sigma/2} \dd z \int \dd \boldsymbol{v} \int \dd \boldsymbol{v}_1 \int_{\Omega_h(z)} \dd \un |\boldsymbol{g} \cdot \widehat{\boldsymbol{\sigma}}|  \overline{f}(\bm{v}_1)\overline{f}(\bm{v}) \\
& (b_{{\widehat{\bm\sigma}}}  -  1 ) \left[\psi\left(\boldsymbol{v}\right)+\psi(\boldsymbol{v}_1)\right].
\end{aligned}
\end{equation}

%% Appendix C %%%%

\section{Decomposition of the Boltzmann collision operator}
\label{apendixc}

  \begin{figure}[t]
\centering
\includegraphics[scale=0.4]{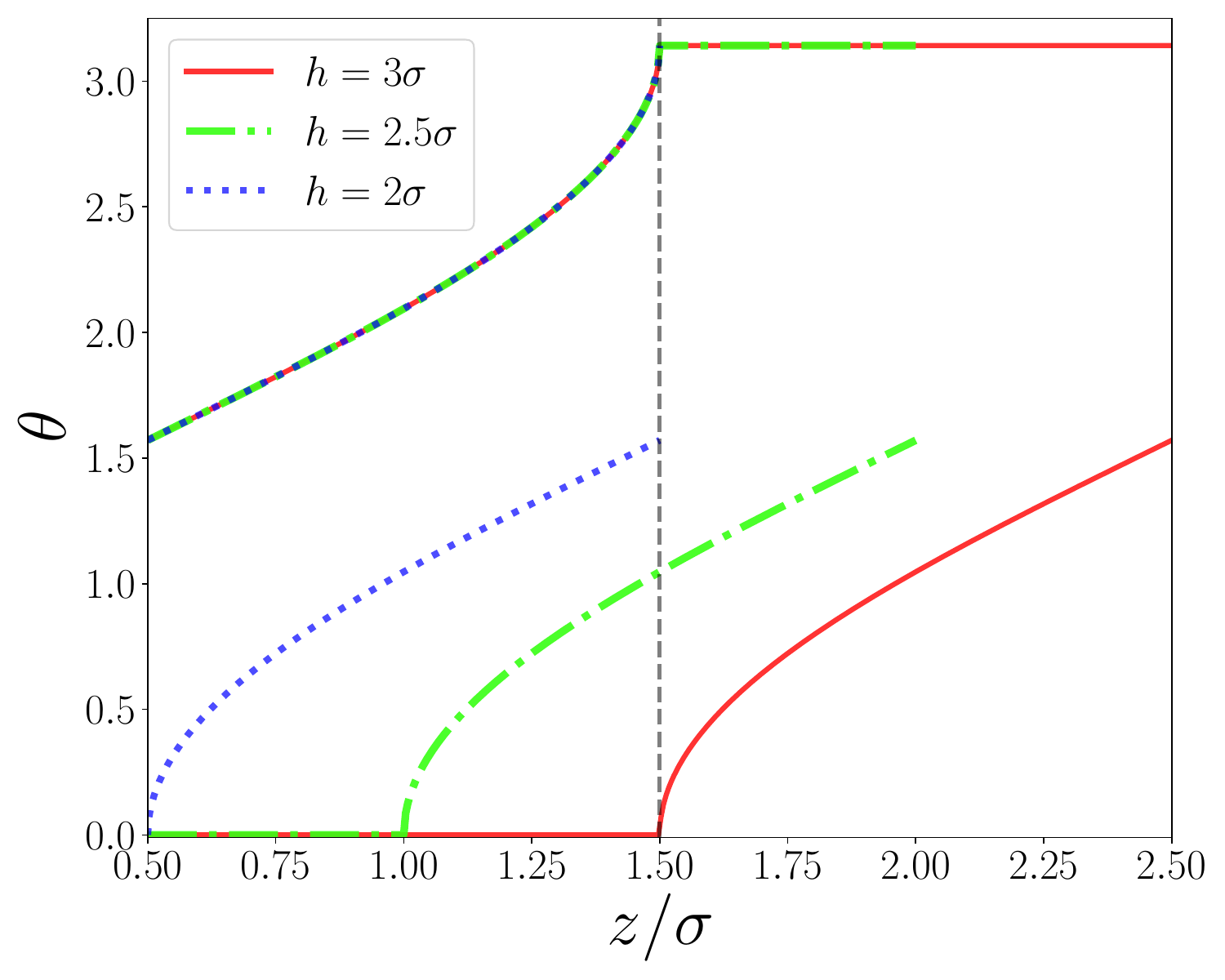}
\caption{ (Color online) $\theta$ as a function of $\frac{z}{\sigma}$ for three different values of $h$. The (red) solid line, (green) dashed--dotted line and (blue) dotted line are the corresponding functions for $h=3 \sigma$, $2.5 \sigma$  and $2\sigma$ respectively. A vertical (black) dashed line is plotted at $\frac{z}{\sigma} = \frac{3}{2} $ to mark the maximum value of $\frac{z}{\sigma}$ for the limit case, $h = 2 \sigma$.      }
\label{fig:inter}
\end{figure}

In this Appendix we obtain  the decomposition given by
Eq.~\eqref{eq:dec_ct} for $ h\geq 2\sigma$. Note that, for these
values, we have a different parametrization of the allowed solid
angles for intermediate heights systems, $2\sigma\leq h \leq 3\sigma$,
and wider systems, $h\geq 3\sigma$. In Fig.~\eqref{fig:inter}, we have
plotted the region of integration concerning the $\theta$ variable as
a function of $\frac{z}{\sigma}$ for three different values of $h$ in
order to visualize their differences. The (red) solid line, (green)
dashed--dotted line and (blue) dotted line are the corresponding
functions for $h=3 \sigma$, $2.5 \sigma$  and $2\sigma$
respectively. A vertical (black) dashed line is plotted at
$\frac{z}{\sigma} = \frac{3}{2} $ to mark the maximum value of
$\frac{z}{\sigma}$ for the limit case, $h = 2 \sigma$.   

Let consider the case for which $h\in [2 \sigma,3 \sigma]$. The
integration domain of Eq.~\eqref{eq:col_op2} can be splitted as  
\begin{equation} \label{eq:c1}
\begin{aligned}
\mathcal{J}_h[\overline{f}|\overline{f}]=&\frac{
  \sigma^{2}}{2(h-\sigma)}  \int_{\sigma / 2}^{h-\sigma / 2} \dd z
\int   \dd \bm{v}_1  \underset{\Omega_h(z) }{ \int} \dd \un|\bm{g}
\vdot \un|(b_{{\widehat{\bm\sigma}}}  -  1 )
\overline{f}\left(\bm{v}_1\right)
\overline{f}\left(\bm{v}\right)\\ 
=& \frac{ \sigma^{2}}{2(h-\sigma)} \int \dd \bm{v}_1  \int_0^{2\pi}
\dd \phi \Bigg\lbrace
\int_{\sigma / 2}^{h-3\sigma / 2} \dd z 
  \int_{0}^{\pi/2}\dd \theta  \sin\theta |\bm{g} \vdot \un|(b_{{\widehat{\bm\sigma}}}  -  1 )
\overline{f}\left(\bm{v}_1\right)
\overline{f}\left(\bm{v}\right)\\
+& \int_{\sigma / 2}^{h-3\sigma / 2} \dd z
\int_{\pi/2}^{\pi/2+b_1(z)} \dd \theta \sin\theta |\bm{g} \vdot 
\un|(b_{{\widehat{\bm\sigma}}}  -  1 ) 
\overline{f}\left(\bm{v}_1\right)
\overline{f}\left(\bm{v}\right)
\\
+& \int_{h-3\sigma / 2}^{3\sigma / 2} \dd z 
  \int_{\pi/2-b_2(z, h)}^{\pi/2}\dd \theta  \sin\theta |\bm{g} \vdot
  \un|(b_{{\widehat{\bm\sigma}}}  -  1 ) 
\overline{f}\left(\bm{v}_1\right)
\overline{f}\left(\bm{v}\right)\\
+& \int_{h-3\sigma / 2}^{3\sigma / 2} \dd z
\int_{\pi/2}^{\pi/2+b_1(z)} \dd \theta \sin\theta |\bm{g} \vdot 
\un|(b_{{\widehat{\bm\sigma}}}  -  1 ) 
\overline{f}\left(\bm{v}_1\right)
\overline{f}\left(\bm{v}\right)
\\
+& \int_{3\sigma / 2}^{h-\sigma / 2} \dd z 
  \int_{\pi/2-b_2(z, h)}^{\pi/2}\dd \theta  \sin\theta |\bm{g} \vdot
  \un|(b_{{\widehat{\bm\sigma}}}  -  1 ) 
\overline{f}\left(\bm{v}_1\right)
\overline{f}\left(\bm{v}\right)\\
+& \int_{3\sigma / 2}^{h-\sigma / 2} \dd z
\int_{\pi/2}^{\pi} \dd \theta \sin\theta |\bm{g} \vdot 
\un|(b_{{\widehat{\bm\sigma}}}  -  1 ) 
\overline{f}\left(\bm{v}_1\right)
\overline{f}\left(\bm{v}\right) \Bigg\rbrace, 
\end{aligned}
\end{equation}
where we have used that  $\theta\equiv [0,\pi/2) \cup (\pi/2,\pi/2+b_1(z)]
 $ for $z\in[\sigma/2,h-3\sigma/2]$, $\theta\equiv
 [\pi/2-b_2(z,h),\pi/2] \cup (\pi/2,\pi]  $ for
 $z\in[3\sigma/2,h-\sigma/2]$, and $\theta\equiv
 [\pi/2-b_2(z,h),\pi/2) \cup (\pi/2,\pi/2+b_1(z)]  $ for
 $z\in(h-3\sigma/2,3\sigma/2)$. As in Appendix \ref{appendixa}, the
 dependence of the distribution function on $\bm{r}_\parallel$ and $t$
 has been omitted. By rearranging terms of 
 Eq.~\eqref{eq:c1}, we get  
\begin{equation} \label{eq:c2}
\begin{aligned}
\mathcal{J}_h[\overline{f}|\overline{f}]=&
\frac{(h-2\sigma)\sigma^2}{2(h-\sigma)}  \int \dd \bm{v}_1
\int_{0}^{2\pi}\dd \phi  \int_{0}^{\pi}\dd \theta \sin\theta |\bm{g}
\vdot \un|(b_{{\widehat{\bm\sigma}}}  -  1 )
\overline{f}\left(\bm{v}_1\right)
\overline{f}\left(\bm{v}, \right) \\ 
&  +\frac{\sigma^2}{2(h-\sigma)} \int \dd \bm{v}_1
\int_{\sigma/2}^{3\sigma/2}  \dd z\int_{0}^{2\pi} \dd \phi
\int_{\pi/2}^{\pi/2+b_1(z)} \dd \theta  \sin\theta |\bm{g} \vdot
\un|(b_{{\widehat{\bm\sigma}}}  -  1 ) \overline{f}\left(
  \bm{v}_1\right) \overline{f}\left(\bm{v}\right)
\\ 
 &+\frac{\sigma^2}{2(h-\sigma)} \int \dd \bm{v}_1
 \int_{h-\sigma/2}^{h-3\sigma/2}  \dd z\int_{0}^{2\pi} \dd \phi
 \int_{\pi/2-b_2(z,h)}^{\pi/2} \dd \theta  \sin\theta |\bm{g} \vdot
 \un|(b_{{\widehat{\bm\sigma}}}  -  1 ) \overline{f}\left(
   \bm{v}_1\right) \overline{f}\left(\bm{v}\right)
 . 
\end{aligned}
\end{equation}
The first term of Eq.~\eqref{eq:c2} can be identified as the ``bulk'' contribution, i.e. 
\begin{align}
\frac{(h-2\sigma)}{(h-\sigma)} \frac{\sigma^2}{2} \int \dd \bm{v}_1
  \int_{0}^{2\pi}\dd \phi  \int_{0}^{\pi}\dd \theta \sin\theta |\bm{g}
  \vdot \un|(b_{{\widehat{\bm\sigma}}}  -  1 )
  \overline{f}\left(\bm{v}_1\right)
  \overline{f}\left(\bm{v}\right) =
  \frac{\ell-1}{\ell} \mathcal{J}^{(b)}[\overline{f}|\overline{f}]. 
\end{align}
By performing the following change of variables, $
\tilde{z} = z-(h-2\sigma)$, in the last term of Eq.~\eqref{eq:c2},
$b_2(z,h)$ transforms into $b_2(\tilde{z},2\sigma)$, so that 
\begin{align}
\begin{aligned}
& \frac{\sigma^2}{2(h-\sigma)} \int \dd \bm{v}_1
\int_{\sigma/2}^{3\sigma/2}  \dd z\int_{0}^{2\pi} \dd \phi
\int_{\pi/2}^{\pi/2+b_1(z)} \dd \theta  \sin\theta |\bm{g} \vdot
\un|(b_{{\widehat{\bm\sigma}}}  -  1 ) \overline{f}\left(
  \bm{v}_1\right) \overline{f}\left(\bm{v}\right)
\\ 
 &+\frac{\sigma^2}{2(h-\sigma)} \int \dd \bm{v}_1
 \int^{3\sigma/2}_{\sigma/2}  \dd \tilde{z}\int_{0}^{2\pi} \dd \phi
 \int_{\pi/2-b_2(z,2\sigma)}^{\pi/2} \dd \theta  \sin\theta |\bm{g}
 \vdot \un|(b_{{\widehat{\bm\sigma}}}  -  1 )
 \overline{f}\left(\bm{v}_1\right)
 \overline{f}\left(\bm{v}\right) \\ 
& = \ell^{-1} \mathcal{J}_{2\sigma}[\overline{f}|\overline{f}].
\end{aligned}
\end{align}  
 Hence, Eq.~\eqref{eq:dec_ct} is obtained. For $h \geq 3\sigma$ the analysis is similar.

%%%%%%%% Apendice F %%%%%%%

\section{Approximate evaluation of $\gamma$}
\label{apendixf}

In this section, the approximate expression of $\gamma$ given by
Eq.~\eqref{eigenvalue1} is evaluated. To obtain the numerator
integral, we consider the B--L collision operator property given by 
Eq.~\eqref{eq:blproperty1} and the isotropy in $v_x,v_y$ plane. So, 
it is obtained
\begin{align} \label{eq:b1}
\begin{aligned}
 \int \dd \bm{v }  \bm{v}_\parallel \cdot \left[\Lambda_h (\bm{v})
   \bm{v}_\parallel  \chi_\text{M} (\bm{v}) \right]  =& \frac{n
   \sigma^{2}}{ (h-\sigma)}\int_{\sigma/2}^{h-\sigma/2} \dd z\int
 \dd \bm{v} \int   \dd \bm{v}_1  \int_{\Omega_h(z)} \dd \un|\bm{g}
 \vdot \un|  \chi_\text{M}(\bm{v}_1) \chi_\text{M}(\bm{v})  v_x \\ 
& \times [b_{\un}^{-1}  - 1  ] v_x .
\end{aligned}
\end{align}
Considering the change of variables
\begin{align}
\begin{aligned}
\bm{g}&= \bm{v}_1 - \bm{v} ,\\
\bm{G}&=\frac{\bm{v}+\bm{v}_1}{2},
\end{aligned}
\end{align}
we get 
\begin{align}\label{eq:int1}
\begin{aligned}
 \int \dd \bm{v }  \bm{v}_\parallel\cdot \left[\Lambda_h (\bm{v})   \bm{v}_\parallel
   \chi_\text{M} (\bm{v}) \right] =&     - \frac{n
   \sigma^{2}}{2(h-\sigma)}\int_{\sigma/2}^{h-\sigma/2} \dd z
 \int_{\Omega_h(z)} \dd \un \int     \dd \bm{g}  \xi(g)  |\bm{g} \vdot
 \un|(\bm{g} \vdot \un)  
   g_x \hat{\sigma}_x ,
   \end{aligned}
\end{align}
where  $\xi(g)  \equiv   (  \frac{m}{4 \pi k_B  T } )^{\frac{3}{2}}  
e^{-\frac{m g^2}{4  k_B T }}$. Expressing the relative velocity,
$\bm{g}$, in the orthonormal base $\left\lbrace \un, \un_{\perp 1} , \un_{\perp 2}
\right\rbrace$ ($ \un_{\perp m} $, $m=1, 2$, are mutually perpendicular
vectors to $\un$), i.e., 
$\bm{g}=g_{\sigma} \un + \sum_{m=1}^{2}g_{\sigma_\perp m} \un_{\perp m
}$, and using the symmetry of the integrand, Eq. (\ref{eq:int1}) can be written in
the form
\begin{align}\label{eq:int2}
\begin{aligned}
 \int \dd \bm{v }  \bm{v}_\parallel \cdot \left[\Lambda_h (\bm{v})  \bm{v}_\parallel \chi_\text{M} (\bm{v}) \right]
=& - \frac{n \sigma^{2}}{2(h-\sigma)} H_1 H_2  ,
   \end{aligned}
\end{align}
where 
\begin{align}
\begin{aligned}
H_1 =& \int     \dd \bm{g}  \xi(g)  |g_\sigma| g_\sigma^2 ,
\end{aligned}
\end{align}
and 
\begin{align}\label{ecH2}
\begin{aligned}
H_2 = &\int_{\sigma/2}^{h-\sigma/2} \dd z \int_{\Omega_h(z)} \dd \un \hat{\sigma}_x^2. 
\end{aligned}
\end{align}
have been introduced. 
$H_1$ can be directly calculated 
\begin{align} \label{eq:solutionh1}
H_1 =&  \left(  \frac{m}{4 \pi k_B  T } \right)^{\frac{3}{2}}  \left(2\int_{0}^{\infty} \dd g_\sigma g_\sigma^3   e^{-\frac{mg_\sigma^2}{4 k_B T }}\right) \left(2\int_{0}^{\infty} \dd g_\perp    e^{-\frac{m g_{\perp}^2}{4 k_B T }}\right)^{2} = \frac{1}{\sqrt{\pi}}\left(\frac{4 k_B T }{m}\right)^{3/2}.
\end{align}
Let us calculate $H_2$ for q2d systems. In this case, the $\un$
integral in Eq. (\ref{ecH2}) is parametrized as
\begin{align}
H_2 = \pi  \int_{\sigma/2}^{h-\sigma/2} \dd z
  \int_{\pi/2-b_1(z)}^{\pi/2+b_2(z,h)} \dd \theta \sin^3 \theta . 
\end{align}
By performing the change of variable
\begin{align}
  \frac{z_1-z}{\sigma} =   \cos  \theta  ,
\end{align} 
$H_2$ can be written in the form
\begin{align} \label{eq:solutionh2}
\begin{aligned}
H_2 =& \frac{\pi}{\sigma}  \int_{\sigma/2}^{h-\sigma/2} \dd z \int^{h-\sigma/2}_{\sigma/2} \dd z_1\left[1- \left(\frac{z_1-z}{\sigma} \right)^2\right]  \\
=& \pi \sigma \ell^2 \left(1-\frac{\ell^2}{6}\right) .
\end{aligned}
\end{align}
 Introducing Eqs. \eqref{eq:solutionh1} and \eqref{eq:solutionh2} into
 \eqref{eq:int2}, we get 
\begin{align}\label{eq:int3}
\begin{aligned}
\int \dd \bm{v }  \bm{v}_\parallel \cdot \left[\Lambda_h (\bm{v})  \bm{v}_\parallel \chi_\text{M} (\bm{v}) \right]
=& - 4  \sqrt{\pi} n \sigma^2   \left(\frac{ k_B T }{m }\right)^{3/2} \ell \left(1-\frac{\ell^2}{6}\right)   .
   \end{aligned}
\end{align}
The denominator of Eq. (\ref{eigenvalue1}) trivially reads
\begin{align}
\begin{aligned}
\int \dd \bm{v } v_\parallel^2  \chi_\text{M} (\bm{v}) &= 2\int \dd \bm{v }  v_x^2  \chi_\text{M} (\bm{v})
 &= \frac{2 k_B T}{m} .
\end{aligned}
\end{align}
Therefore, for q2d systems, the eigenvalue is  
 \begin{align}
 \gamma   =  -   \frac{8\sqrt{\pi} n \sigma^2}{3}    \left(\frac{ k_B T }{ m  }\right)^{1/2} \frac{\ell\left(6-\ell^2\right)}{8} .
 \end{align}

%%%%%%%%%%%%%%%%%%%%%%%%%

For $h\geq 2 \sigma$, using the decomposition given by Eq.~\eqref{eq:bloperator_divided}, we have 
\begin{align}\label{eq:eigen_wide_system}
\gamma = \frac{\ell-1}{\ell}\frac{  \int \dd \bm{v }  \bm{v}_\parallel \cdot \left[\Lambda^{(b)} (\bm{v})  \bm{v}_\parallel \chi_\text{M} (\bm{v}) \right]}{\int \dd \bm{v } v_\parallel^2  \chi_\text{M} (\bm{v})}+\frac{1}{\ell}\frac{  \int \dd \bm{v }  \bm{v}_\parallel \cdot \left[\Lambda_{2\sigma} (\bm{v})  \bm{v}_\parallel \chi_\text{M} (\bm{v}) \right]}{\int \dd \bm{v } v_\parallel^2  \chi_\text{M} (\bm{v})}.
\end{align} 
 The first integral is easy to compute, since there are not solid angles restrictions
\begin{align}\label{eq:int_1_wide_system}
\frac{  \int \dd \bm{v }  \bm{v}_\parallel \cdot \left[\Lambda^{(b)} (\bm{v})  \bm{v}_\parallel \chi_\text{M} (\bm{v}) \right]}{\int \dd \bm{v } v_\parallel^2  \chi_\text{M} (\bm{v})} =  -\frac{8    \sqrt{\pi} n \sigma^2}{3} \left(\frac{k_B T}{m} \right)^{1/2}. 
\end{align}
The second integral is just proportional to the $\gamma$ calculated previously for $h=2\sigma$
\begin{align}\label{eq:int_2_wide_system}
\frac{  \int \dd \bm{v }  \bm{v}_\parallel \cdot \left[\Lambda_{2\sigma} (\bm{v})  \bm{v}_\parallel \chi_\text{M} (\bm{v}) \right]}{\int \dd \bm{v } v_\parallel^2  \chi_\text{M} (\bm{v})} =  -\frac{5  n  \sqrt{\pi} \sigma^2}{3} \left(\frac{k_B T}{m} \right)^{1/2}. 
\end{align} 
Introducing Eqs. \eqref{eq:int_1_wide_system} and
\eqref{eq:int_2_wide_system}  into \eqref{eq:eigen_wide_system}, we
finally have 
\begin{align}
\gamma  = - \frac{8  \sqrt{\pi} n \sigma^2}{3 } \left(\frac{k_B T}{m} \right)^{1/2} \frac{(8\ell-3)}{8 \ell} .
\end{align}

\newpage

%%%%%%%%%%%%%% BIBLIOGRAPHY


\begin{thebibliography}{10}
\bibitem{bgm24erratum} Note that there are two misprints in Eq.~(74) of
  \cite{bgm24}: the $\alpha$ term should be divided by $h-\sigma$ and
  the collision term as well (instead of $h-z$).
\end{thebibliography}
\end{document}